\documentclass[11pt]{article} 

\usepackage{a4wide}
\usepackage{amssymb}
\usepackage{amsmath}
\usepackage{color} 
\usepackage{epic} 
\usepackage{here}

\newcommand{\nco}{\newcommand}
\newfont{\msbm}{msbm10 at 12pt}
\nco{\CC}{\mbox{\msbm C}} 
\nco{\ZZ}{\mbox{\msbm Z}}
\nco{\NN}{\mbox{\msbm N}}
\nco{\II}{\mbox{\msbm I}} 
\nco{\RR}{\mbox{\msbm R}}
\nco{\munite}{\ensuremath{\,\,\mathrm{l}\!\!\!1}} 
\nco{\dps}{\displaystyle}
\nco{\ov}{\overline}
\nco{\ud}{\underline}

\newtheorem{prop}{Property} 
\newtheorem{theo}{Theorem} 
\newtheorem{defin}{Definition}

\title{Higher Coxeter graphs associated to affine $su(3)$ modular invariants
\vspace{1.0cm}}

\author{{\bf D. Hammaoui}${}^{1}$\thanks{E-mail:d.hammaoui@sciences.univ-oujda.ac.ma}, {\bf G. Schieber}${}^{1,2}$\thanks{Supported by a fellowship of FAPERJ - Funda\c{c}\~ao de Amparo \`a Pesquisa do Estado do Rio de Janeiro, Brasil. E-mail: schieber@cbpf.br}\;, {\bf E. H. Tahri}${}^{1}$\thanks{E-mail:tahrie@sciences.univ-oujda.ac.ma}.  \\
\\
${}^1$ {\it Laboratoire de Physique Th\'eorique et des Particules (LPTP)} \\ 
{\it D\'epartement de Physique, Facult\'e des Sciences} \\
                 {\it Universit\'e Mohamed I, B.P.524 Oujda 60000, Maroc}           \\
\\
${}^2$ {\it CBPF - Centro Brasileiro de Pesquisas F\'{\i}sicas} \\
                 {\it Rua Dr. Xavier Sigaud, 150}\\
                 {\it 22290-180, Rio de Janeiro, Brasil}\\
}

\date{}
\begin{document}
\thispagestyle{empty}

\begin{titlepage}

\maketitle

\vspace{1cm}

\begin{abstract}
The affine $su(3)$ modular invariant partition functions in 2d RCFT are associated with a set of generalized Coxeter graphs. These partition functions fall into two classes, the block-diagonal (Type I) and the non block-diagonal (Type II) cases, associated, from spectral properties, to the subsets of subgroup and module graphs respectively. 
We introduce a modular operator $\hat{T}$ taking values on the set of vertices of the subgroup graphs. It allows us to obtain easily the associated Type I partition functions. We also show that all Type II partition functions are obtained by the action of suitable twists $\vartheta$ on the set of vertices of the subgroup graphs. These twists have to preserve the values of the modular operator $\hat{T}$.
\end{abstract}

\vfill




\vspace*{0.5 cm}

\end{titlepage}


\section{Introduction}

The classification of 2d modular invariant partition functions of affine $su(3)$ WZWN model has been established in \cite{gannon-class}. These partition functions fall into two classes. In the first one, called Type I, the partition functions are block-diagonal with respect to an ambichiral\footnote{Some authors call it the extended algebra.} algebra $\mathcal{B}$. Let $\chi_{\lambda}$ be the affine characters of $su(3)$, the extended characters 
$\hat{\chi}_c$ are defined as linear combination of the $\chi_{\lambda}$'s: 
$\hat{\chi}_c = \sum_{\lambda} b_{c , \lambda} \, \chi_{\lambda}$. The modular invariant partition function reads $\mathcal{Z}=\sum_{c \in \mathcal{B}} |\hat{\chi_c}|^2$. In the second class, called Type II, the partition functions are given by $\mathcal{Z}=\sum_{c \in \mathcal{B}} \hat{\chi_c} \, \overline{\hat{\chi}_{\vartheta(c)}}$, where $\vartheta(c)$ is a non-trivial twist of elements of $\mathcal{B}$. We show in this paper, from the association of modular invariants with graphs,  that the ambichiral algebra $\mathcal{B}$ and the twists $\vartheta$ can be characterized by the values of a modular operator $\hat{T}$.\\

The so-called $ADE$ classification of affine $su(2)$ modular invariants \cite{CIZ-class} bears its name from the fact 
that spectral properties of these Dynkin diagrams (Coxeter exponents) are encoded in the diagonal elements of the modular invariants. The affine $su(3)$ classification has also been related to a set of (generalized) Coxeter graphs, whose empirical list was proposed from similar spectral properties \cite{DiF_Zub-graphs,DiF_Zub-2,Pet_Zub-CFT}. Later on, in a rather different approach, the Ocneanu construction associated to every $ADE$ Dynkin diagram a special kind of finite dimensional weak Hopf algebra (WHA) \cite{Oc-paths} (see also \cite{Pet_Zub-Oc,Gil-tesis,Coq_Trinchero}). 
In this construction, vertices of the $A$ graphs are labelled by irreps of $SU(2)_q$ at roots of unity, the quantum version of irreps of $SU(2)$; while vertices of the other graphs are labelled by irreps of ``modules'' or ``subgroups'' of 
$SU(2)_q$ (the quantum McKay correspondance \cite{qMcKay}). It is generalizing these notions to the $su(3)$ case that 
the set of generalized Coxeter graphs proposed in \cite{DiF_Zub-graphs} has been slightly amended in \cite{Oc-Bariloche}, where the final list is published.  They can be seen as labelling ``modules'' or ``subgroups'' of 
$SU(3)_q$. For the $su(2)$ model, subgroup graphs are $A_n, D_{2n}, E_6$ and $E_8$, while module graphs are $D_{2n+1}$ and $E_7$. For the $su(3)$ model, we have the $\mathcal{A}$ series, which are truncated Weyl alcoves, at level $k$, of $su(3)$. There are two operations defined on these graphs, the conjugation and the $\mathbb{Z}_3$-symmetry. The entire list of $SU(3)$-type graphs consists of 4 infinite series: the $\mathcal{A}_k$ series and its conjugated $\mathcal{A}_k^*$, the orbifold $\mathcal{D}_k=\mathcal{A}_k/3$ series and its conjugated $\mathcal{D}_k^*$; and 7 exceptional graphs $\mathcal{E}_5,\mathcal{E}_5^*=\mathcal{E}_5/3,\mathcal{E}_9,\mathcal{E}_9^*=\mathcal{E}_9/3,\mathcal{E}_{21},\mathcal{D}_9^{\xi}$ and $\mathcal{D}_9^{\xi*}$. The subset of subgroup graphs is the infinite series $\mathcal{A}_k$ (for all $k$) and $\mathcal{D}_k$ (for $k=0\mod3$), and the three exceptional graphs $\mathcal{E}_5,\mathcal{E}_9$ and $\mathcal{E}_{21}$. \\

In this paper, we show how to obtain all modular invariant partition functions from the subset of subgroup graphs. We determine the coefficients $b_{\lambda,c}$ from induction-restriction maps, and the ambichiral algebra $\mathcal{B}$ and the twists $\vartheta$ from the action of a modular operator $\hat{T}$ on the vertices of the subgroup graphs. We therefore do not discuss the module graphs. They are defined from well-known conjugation (see \cite{DiF_Zub-graphs, Evans-conj, Pet_Zub-conj}) and orbifold (see \cite{Kostov,Fendley}) methods from the subgroup graphs. The subgroup graphs are taken here as initial data. The infinite series $\mathcal{A}_k$ and its orbifold series $\mathcal{D}_k=\mathcal{A}_k/3$ for $k=0\mod3$ are constructed from primary data. The exceptional subgroup graphs $\mathcal{E}_5,\mathcal{E}_9$ and $\mathcal{E}_{21}$ are taken from \cite{DiF_Zub-graphs} or \cite{Oc-Bariloche}.
We are not aware of any written formulae for the explicit construction of the WHA associated with a generalized Coxeter graphs. Nevertheless, its structure maps have to satisfy well-defined equations \cite{Oc-Bariloche, Pet_Zub-Oc,Gil-tesis}. One interesting structure of the WHA is the algebra of quantum symmetries. The expressions for Type I and Type II modular invariant partition functions can be shown to come from the bimodule structure of the algebra of quantum symmetries under the fusion algebra \cite{Gil-tesis}, which furthermore allows the determination of more general partition functions, corresponding to systems on a torus with two defect lines \cite{Pet_Zub-gener}.
It is not the purpose of this article to study the algebra of quantum symmetries of the generalized Coxeter graphs.
But let us mention that Type II partition functions can be obtained from a special twist of the Type I ones is due to the fact that quantum symmetries of module graphs are constructed from a twist of the quantum symmetries of a subgroup graph, called its parent graph (see for example \cite{Pet_Zub-Oc,Coq_Gil-Tmod, Gil-tesis}). The modular operator $\hat{T}$ defined in this paper is a usefull tool for a realization of the algebra of quantum symmetries (see \cite{Coq_Gil-ADE,Coq_Gil-Tmod,Gil-tesis}), whose study will be presented in a forthcoming article \cite{Qsymsu3}.\\

The method presented in this article is the following. Vertices of the $\mathcal{A}_k$ graph are labelled by irreps $\hat{\lambda}$ of 
$\widehat{su}(3)_k$. The modular matrix $\mathcal{T}$ -- one of the two generators of the modular group $PSL(2,\mathbb{Z})$ -- is a diagonal matrix that acts on the characters of an irrep $\hat{\lambda}$ of $\widehat{su}(3)_k$.
This allows us to define a value, called modular operator value $\hat{T}$, on each vertex of the $\mathcal{A}_k$ graph. 
We call $G$ a subgroup graph. The vector space spanned by the vertices of the graph $G$ is a module under the action of the graph algebra of the graph $\mathcal{A}_k$ with same norm. We can therefore define branching rules $\mathcal{A}_k \hookrightarrow G$. For $\lambda \in \mathcal{A}_k$ and $c \in G$, we have:
$\lambda \hookrightarrow \sum_{c} b_{\lambda,c} \,c$. The branching coefficients $b_{\lambda,c}$ can be encoded in a matrix $E_0$, called essential matrix or intertwiner. From the induction-restriction maps between 
$\mathcal{A}_k$ and $G$, one can try to define a modular operator value on vertices of $G$. This is only possible for a subset $\mathcal{B}$ of vertices of $G$. This subset $\mathcal{B}$ is an algebra, called the ambichiral algebra (a subalgebra of the graph algebra of $G$). Type I partition functions can then easily be obtained from the 
previous block-diagonal relations. We then analyse all involutions $\vartheta$ that one can define on the subset of vertices of $G$ belonging to $\mathcal{B}$, such that they leave invariant the modular operator $\hat{T}$. We obtain in this way all Type II modular invariant partition functions. To our knowledge, restricting the association of all affine $su(3)$ modular invariants to the subset of subgroup graphs and twists $\vartheta$ characterized by the modular operator $\hat{T}$ is new. \\

Finally, let us notice that several constructions used here can be understood in terms of planar algebras \cite{Jones}, modular tensor categories \cite{Fuchs1}, or in particular in terms of nets of subfactors \cite{be,bek,Evans,Evans-2}, but we shall not use these notions. \\ 

The article is organized as follows. Section {\bf 2} is a short review on modular invariant partition functions in 2d RCFT. Section {\bf 3} treats explicitely the affine $su(3)$ model.
We define the conjugation, the $\mathbb{Z}_3$-symmetry and the modular operator $\hat{T}$ on irreps of affine $su(3)$. In Section {\bf 4} are introduced the generalized Coxeter graphs associated to the $su(3)$ classification. We define the induction-restriction mechanism and the modular properties of the subset of subgroup graphs. 
The explicit analysis of the different cases is made in the last three sections. 
In Section {\bf 5} we treat the $\mathcal{A}_k$ series. It leads to diagonal modular invariants. Three types of twists $\vartheta$ preserving $\hat{T}$ can be defined, namely conjugation, a twist $\rho$ related to the $\mathcal{Z}_3$ symmetry and the combination of conjugation and $\rho$. We obtain the Type II modular invariant series $\mathcal{A}_k^*$ (for all $k$) and $\mathcal{D}_k=\mathcal{A}_k/3$ and $\mathcal{D}_{k}^*$ (for $k \not=0\mod3$). In Section 
{\bf 6} we describe the $\mathbb{Z}_3$-orbifold procedure defining the graphs $\mathcal{D}_k=\mathcal{A}_k/3$. These graphs have a fusion graph algebra for $k=0 \mod 3$ (subgroup case).
The $\hat{T}$-invariant twist is the conjugation, leading to the $\mathcal{D}_k^*$ series. There is also an exceptional $\hat{T}$-invariant twist $\xi$ at level $k=9$, together with its conjugate. The modular invariants obtained in these two sections are sometimes denoted by $\mathcal{ADE}_7$ in the litterature \cite{ganon-moddata}. The three exceptional subgroup graphs $\mathcal{E}_5$, $\mathcal{E}_9$ and 
$\mathcal{E}_{21}$ and their $\hat{T}$-invariant conjugation and twist are treated in Section {\bf 7}. We finish with concluding remarks and open questions. The correspondance between the affine $su(3)$ classification and the generalized Coxeter graphs is presented in the appendix.


\section{RCFT data and partition functions}
Rational Conformal Field Theory (RCFT) in 2d can be conventionally described by a set of data of different nature. Considering only the chiral half of the theory (only holomorphic or anti-holomorphic variables), the chiral set of data is: the chiral algebra $\mathfrak{g}$, its finite set $\mathcal{I}$ of irreducible representations $\{ {\mathcal V}_i \}_{i \in \mathcal{I}}$ and the characters $\chi_i(q)$ of a given representation ${\mathcal V}_i$, where 
$q = \exp (2 i \pi / \tau)$, with $\tau$ a complex number in the upper half-plane.

For the WZWN models the chiral algebra 
$\mathfrak{g}$ is an affine Lie algebra $\widehat{g}_k$ of rank $r$ at level $k$ (Kac-Moody algebra). Integral highest weight representations $\hat{\lambda}$ (that we will call irreps) of 
$\widehat{g}_k$ are labelled by $\hat{\lambda} = (\lambda_0, \lambda_1, \ldots, \lambda_r)$, where $\lambda_i$ are the Dynkin labels. In a level $k$ the number of irreps is 
finite and we denote its set by $\mathcal{P}_+^{k}$. The $\lambda_0$ label is fixed once the level is known, so we write $\lambda = (\lambda_1, \ldots, \lambda_r)$
for $\hat{\lambda}$. The normalized characters $\chi_{\lambda}^k$ in an irrep $\lambda$ of 
$\widehat{g}_k$ have a simple transformation property with respect to the modular group 
$PSL(2,\ZZ)$, generated by the two transformations $\mathcal{S}$ and $\mathcal{T}$:
\begin{equation}
\mathcal{T}: \tau \mapsto \tau + 1 \;, \qquad \qquad \qquad  
\mathcal{S}: \tau \mapsto - \frac{1}{\tau} \;,
\end{equation}
satisfying the relations $\mathcal{S}^2 = (\mathcal{S}\mathcal{T})^3 = \mathcal{C}$, with $\mathcal{C}^2 = \munite$. These characters define a unitary representation of the modular group:
\begin{equation}
\sum_{\mu \in \mathcal{P}^{k}_+} \mathcal{T}_{\lambda \mu} \; \chi_{\mu}^k (\tau) 
= \chi_{\lambda}^k (\tau +1)\;, \qquad \qquad \qquad 
\sum_{\mu \in \mathcal{P}^{k}_+} \mathcal{S}_{\lambda \mu} \; \chi_{\mu}^k (\tau) 
= \chi_{\lambda}^k (-1/\tau) \;.
\end{equation}
Expression for the modular matrices $\mathcal{S}_{\lambda \mu}$ and 
$\mathcal{T}_{\lambda \mu}$ for any Kac-Moody algebra $\widehat{g}_k$ at level $k$ may be found in \cite{FMS-book,kac}. The fusion of representations of chiral algebras in RCFT reads:
\begin{equation}
\lambda \otimes \mu = \sum_{\nu \in \mathcal{P}^{k}_+} \mathcal{N}_{\lambda \mu}^{\nu} \; \nu
\; , \qquad \qquad \mathcal{N}_{\lambda \mu}^{\nu} \in \NN \;,
\end{equation}
where the multiplicities $\mathcal{N}_{\lambda \mu}^{\nu}$ are called the fusion coefficients. They can be obtained from the elements of the modular $\mathcal{S}$ matrix through the Verlinde formula \cite{verlinde}:
\begin{equation}
{\mathcal N}_{\lambda \mu}^{\nu} = \sum_{\beta \in \mathcal{P}_{+}^{k}} 
\frac{S_{\lambda \beta} \, S_{\mu \beta } \, S_{\nu \beta}^{*}}{S_{0 \beta}} \; ,
\end{equation}
where $\lambda =0$ is the trivial representation. 

A physically sensible conformal theory is defined on a Hilbert space $\mathcal{H}$ which is constructed combining the holomorphic and anti-holomorphic sectors of the theory:
\begin{equation}
{\mathcal H} = \bigoplus_{\lambda, \mu \in \mathcal{P}_{+}^{k}} {\mathcal M}_{\lambda\mu} \; \lambda \otimes \mu^{*} \; ,
\end{equation}
where ${\mathcal M}_{\lambda \mu}$ specifies the multiplicity of 
$\lambda \otimes \mu^{*}$ on $\mathcal{H}$ ($\mu^*$ is the conjugated irrep of $\mu$). The $\widehat{g}_k$ WZWN partition function is given by:
\begin{equation}
{\mathcal Z} (\tau) = \sum_{\lambda, \mu \in \mathcal{P}_{+}^{k}} {\mathcal M}_{\lambda\mu} \; \chi_{\lambda} (\tau) \; \overline{\chi_{\mu} (\tau)} \; .
\end{equation}
$\mathcal{Z}$ is modular invariant if the $|\mathcal{I}| \times |\mathcal{I}|$ matrix $\mathcal{M}_{\lambda\mu}$ satisfies the following 3 properties:
\begin{itemize}  
\item[(P1)] $\mathcal{M}_{\lambda\mu} \in \NN$\,.
\item[(P2)] $\mathcal{M}_{00} =1$, where $\lambda = 0$ is the trivial representation (unicity of the vacuum).
\item[(P3)] $\mathcal{M}$ commutes with the generators $\mathcal{S}$ and $\mathcal{T}$ of the modular group (modular invariance).
\end{itemize}

The search of modular invariant partition functions is reduced to the classification of the matrices $\mathcal{M}_{\lambda\mu}$ satisfying properties (P1)-(P2)-(P3).


\section{The affine $su(3)$ case}
We label integral highest weight representations $\hat{\lambda}$ of $\widehat{su}(3)_k$ by their finite part $\lambda = (\lambda_1,\lambda_2)$. Its set $\mathcal{P}^k_+$ is given by: 
\begin{equation}
\mathcal{P}^k_+ = \{ \lambda = (\lambda_1, \lambda_2)\, | \, \lambda_1, \lambda_2 \in 
\mathbb{N} \, , \, \lambda_1 + \lambda_2 \leq k \} \;,
\end{equation}
of cardinality $(k+1)(k+2)/2$. The trivial representation is $\lambda = 0 = (0,0)$ and the two fundamental irreps are $(1,0)$ and $(0,1)$. Irreps
shifted by $\rho = (1,1)$ are denoted by $\lambda^{\prime}$, so that $\lambda^{\prime}
=(\lambda_1+1, \lambda_2+1)$. 
The altitude $\kappa$ is defined by $\kappa = k+3$. It will also be called generalized Coxeter number. 
The explicit values of the modular matrices $\mathcal{S}$ and $\mathcal{T}$ for 
$\widehat{su}(3)_{k}$, using the shifted irreps $\lambda^{\prime }$, are given
by:
\begin{eqnarray}
\mathcal{T}_{\lambda \mu } &=&\exp \left[ 2i\pi \left( \frac{\lambda
_{1}^{^{\prime }2}+\lambda _{1}^{^{\prime }}\lambda _{2}^{^{\prime
}}+\lambda _{2}^{^{\prime }2}}{3\kappa }-\frac{1}{3}\right) \right] \delta
_{\lambda \mu }=e_{\kappa }\left[ \kappa -(\lambda _{1}^{^{\prime
}2}+\lambda _{1}^{\prime }\lambda _{2}^{\prime }+\lambda _{2}^{^{\prime }2}) \right]  \delta_{\lambda\mu}
\label{Tmodmatr} \\
\mathcal{S}_{\lambda \mu } &=&\frac{-i}{\sqrt{3}\kappa }\left\{ e_{\kappa
}[2\lambda _{1}^{^{\prime }}\mu _{1}^{^{\prime }}+\lambda _{1}^{^{\prime
}}\mu _{2}^{^{\prime }}+\lambda _{2}^{^{\prime }}\mu _{1}^{^{\prime
}}+2\lambda _{2}^{^{\prime }}\mu _{2}^{^{\prime }}]-e_{\kappa }[-\lambda
_{1}^{^{\prime }}\mu _{1}^{^{\prime }}+\lambda _{1}^{^{\prime }}\mu
_{2}^{^{\prime }}+\lambda _{2}^{^{\prime }}\mu _{1}^{^{\prime }}+2\lambda
_{2}^{^{\prime }}\mu _{2}^{^{\prime }}]\right.  \notag \\
&&\qquad -e_{\kappa }[2\lambda _{1}^{^{\prime }}\mu _{1}^{^{\prime
}}+\lambda _{1}^{^{\prime }}\mu _{2}^{^{\prime }}+\lambda _{2}^{^{\prime
}}\mu _{1}^{^{\prime }}-\lambda _{2}^{^{\prime }}\mu _{2}^{^{\prime
}}]+e_{\kappa }[-\lambda _{1}^{^{\prime }}\mu _{1}^{^{\prime }}+\lambda
_{1}^{^{\prime }}\mu _{2}^{^{\prime }}-2\lambda _{2}^{^{\prime }}\mu
_{1}^{^{\prime }}-\lambda _{2}^{^{\prime }}\mu _{2}^{^{\prime }}]  \notag \\
&&\qquad +e_{\kappa }[-\lambda _{1}^{^{\prime }}\mu _{1}^{^{\prime
}}-2\lambda _{1}^{^{\prime }}\mu _{2}^{^{\prime }}+\lambda _{2}^{^{\prime
}}\mu _{1}^{^{\prime }}-\lambda _{2}^{^{\prime }}\mu _{2}^{^{\prime
}}]-e_{\kappa }[-\lambda _{1}^{^{\prime }}\mu _{1}^{^{\prime }}-2\lambda
_{1}^{^{\prime }}\mu _{2}^{^{\prime }}-2\lambda _{2}^{^{\prime }}\mu
_{1}^{^{\prime }}-\lambda _{2}^{^{\prime }}\mu _{2}^{^{\prime }}]  \notag \\
&& \label{Smodmatr}
\end{eqnarray}
where $e_{\kappa }[x]:=\exp [\frac{-2i\pi x}{3\kappa }]$. 
The 6 outer automorphisms of the ring of $su(3)$ irreps are generated by $C$ (order
2) and $A$ (order 3). C is the conjugation automorphism (corresponding to
charge conjungation in CFT):
\begin{equation}
C(\lambda _{1},\lambda _{2})=(\lambda _{2},\lambda _{1})\;,
\end{equation}
that we will also write sometimes $C(\lambda)= \lambda^*$. 
$A$ is the $\ZZ_{3}$ automorphism:
\begin{equation}
A(\lambda _{1},\lambda _{2})=(k-\lambda _{1}-\lambda _{2},\lambda _{1})\;.
\label{z3sym}
\end{equation}
Notice that $A^{2}(\lambda _{1},\lambda _{2})=(\lambda_{2},k-\lambda
_{1}-\lambda _{2})$ and $A^{3}=\munite$. The triality $t(\lambda )$ is defined by:
\begin{equation}
t(\lambda _{1},\lambda _{2})=\lambda _{1}-\lambda _{2}\qquad \qquad \mod3\;.
\end{equation}
The modular $\mathcal{T}$ matrix (\ref{Tmodmatr}) being diagonal, each irrep $\lambda$ of
$\widehat{su}(3)_k$ carries a well-defined value of $\mathcal{T}$, that we call the modular operator value $\hat{T}$ of $\lambda$.  
\begin{defin}
The modular operator value $\hat{T}$ of an irrep $\lambda = (\lambda_1,\lambda_2)$ is defined by:
\begin{equation}
\hat{T}[\lambda] =  (\lambda _1 +1)^{2} + (\lambda_1 + 1)(\lambda_2 + 1)
+ (\lambda_2 + 1)^2 \quad \mod 3\kappa \;.
\label{Tmodop}
\end{equation}  
\end{defin}
An important property, derived from Equation (\ref{Tmodmatr}), is that the modular invariant $\mathcal{M}$ commutes with the modular matrix $\mathcal{T}$ iff the following condition is satisfied \cite{gannon-class}:
\begin{equation}
\hat{T} [\lambda] = \hat{T}  [\mu] \;, \qquad \qquad \qquad \textrm{ whenever } 
\mathcal{M}_{\lambda \mu} \not= 0 \;.
\end{equation}
\begin{defin}
The twist operator $\rho$ acting on an irrep $\lambda = (\lambda _1, 
\lambda _2)$ is defined by:
\begin{equation}
\rho (\lambda) = A^{k t(\lambda)} (\lambda) \;,
\end{equation}
where $k$ is the level and $t(\lambda)$ is the triality of $\lambda$. 
\end{defin}
A straightforward calculation shows that $\rho$ is an involution, i.e. 
$\rho^2(\lambda)=\lambda$. In fact, like the conjugation, it is a $\hat{T}$-invariant involution:
\begin{prop}
The modular operator $\hat{T}$ has the same values in an irrep $\lambda$, its conjugated $C(\lambda)=\lambda^*$ and ``its twisted'' $\rho(\lambda)$:
\begin{equation}
\hat{T}[\lambda] = \hat{T}[\lambda^*] = \hat{T}[\rho(\lambda)] \;.
\label{proprho}
\end{equation} 
\end{prop}
\textbf{Proof} The conjugated case is trivial from the definition of the modular operator. The triality of an irrep $\lambda $ is $t(\lambda )\in \{0,1,2\}$:\\
i- For $kt(\lambda )=0\mod3$ (i.e. $k=0\mod3$ or $t(\lambda )=0\mod3)$,
because $A^{3}=\munite$, we have $A^{kt(\lambda )}=\munite$, so $\rho
(\lambda )=\lambda $. \\
ii- For $kt(\lambda )=1\mod3$ (i.e. $k=1\mod3$ and $t(\lambda )=1\mod3$, or 
$k=2 \mod3$ and $t(\lambda )=2\mod3$), we have $\rho (\lambda )=A(\lambda )$ , a
straightforward computation gives $\hat{T}[A(\lambda )]=\hat{T}[\lambda ] - 
(k-t(\lambda )-3\lambda_{2})(k+3) = \hat{T}[\lambda ]+ 3(k+3)\lambda_{2} = 
\hat{T}[\lambda]$ because $k - t(\lambda ) = 0 \mod3 $, $\lambda_{2}\in \mathbb{N}$, and 
$\hat{T}[\lambda ]$ is defined $\mod3(k+3)$. \\
iii- For $kt(\lambda )=2\mod3$ (i.e. $k=1\mod3$ and $t(\lambda )=2\mod3$, or $k=2
\mod3$ and $t(\lambda )=1\mod3$), we have $\rho (\lambda )=A^{2}(\lambda )$,
and we find $\hat{T}[A^{2}(\lambda )]=\hat{T}[\lambda]-(k-$ $2t(\lambda )-3\lambda
_{2})(k+3)=\hat{T}[\lambda] + 3(k+3)\lambda_{2} = \hat{T}[\lambda]$ because 
$k-2t(\lambda )=0\mod3$. \\
So Equation (\ref{proprho}) is verified in all cases. \hfill $\blacksquare $ \\

The affine $su(3)$ classification obtained in \cite{gannon-class} consists of 4 infinite series
(for all levels) and 6 exceptional cases (at levels $k=5,9$ and $21$). To this classification has been associated a set of generalized Coxeter graphs, called the Di Francesco-Zuber graphs (see next sections).
Notice that two different graphs can be associated with the same partition function.
Using the conventions adopted in this paper, the affine $su(3)$ classification and its associated graphs are presented in the appendix.


\section{Classifications, nimreps and graphs}
We present in this section the relation between the classification of affine $su(3)$ partition functions and {\sl nimreps} (``numerical integer valued matrix representations'') of algebra structures and graphs.

\subsection{Fusion matrices $N_{\lambda}$ and $\mathcal{A}_k$ graphs}
Consider the finite set of irreps $\lambda \in \mathcal{P}_+^{k}$ of $\widehat{su}(3)_k$, of cardinality $|\mathcal{I}|=(k+1)(k+2)/2$. The fusion of irreps of 
$\widehat{su}(3)_k$ is given by:
\begin{equation}
\lambda \otimes \mu = \sum_{\nu \in \mathcal{P}_+^k} \mathcal{N}_{\lambda\mu}^{\nu} \; \nu \;,
\end{equation}  
where the fusion coefficients $\mathcal{N}_{\lambda\mu}^{\nu}$ can be calculated by the Verlinde formula using the modular $\mathcal{S}$ matrix (\ref{Smodmatr}), but are also given by simple recurrence relations (see below). The {\bf fusion algebra} is a commutative and associative algebra, with generators $g_{\lambda}, \lambda=0, \ldots, |\mathcal{I}|-1$, unity $g_0$ 
and structure coefficients given by the fusion coefficients: $g_{\lambda} \, g_{\mu} = \sum_{\nu}\mathcal{N}_{\lambda\mu}^{\nu} \, g_{\nu}$. The fusion matrices $N_{\lambda}$ are
$|\mathcal{I}| \times |\mathcal{I}|$ matrices defined by $(N_{\lambda})_{\mu\nu} = \mathcal{N}_{\lambda\mu}^{\nu}$.
They form a faithfull representation of the fusion algebra:
\begin{equation}
N_{\lambda} \, N_{\mu} = \sum_{\nu \in \mathcal{P}_+^k} \mathcal{N}_{\lambda\mu}^{\nu} \, N_{\nu} \;.
\end{equation}
From the properties of the fusion algebra and of the $\mathcal{S}$ matrix, the fusion coefficients 
$\mathcal{N}_{\lambda\mu}^{\nu}$ satisfy the following properties:
\begin{equation}
\mathcal{N}_{\lambda\mu}^{\nu}  = \mathcal{N}_{\mu \lambda}^{\nu}  = \mathcal{N}_{\lambda^*\mu^*}^{\nu^*} = 
\mathcal{N}_{\lambda^*\nu}^{\mu} \; .
\end{equation}
For the trivial representation $\lambda=0=(0,0)$ we have $N_{(0,0)}=\munite$. There are two fundamental irreps $(1,0)$ and $(0,1)=(1,0)^*$. The fusion matrix $N_{(0,1)}$ is the transposed matrix of $N_{(1,0)}$. Once $N_{(1,0)}$ is known, the other fusion matrices can be obtained from the {\bf truncated recursion formulae of $SU(3)$ irreps}, applied for increasing level up to $k$:
\begin{eqnarray}
N_{(\lambda,\mu)} &=& N_{(1,0)} \, N_{(\lambda-1,\mu)} - N_{(\lambda-1,\mu-1)} - 
N_{(\lambda-2,\mu+1)} \qquad \qquad \textrm{if } \mu \not= 0 \nonumber \\
N_{(\lambda,0)} &=& N_{(1,0)} \, N_{(\lambda-1,0)} - N_{(\lambda-2,1)} 
\label{su3recform} \\
N_{(0,\lambda)} &=& (N_{(\lambda,0)})^{tr} \nonumber
\end{eqnarray} 
where it is understood that $N_{(\lambda,\mu)} =0 $ if $\lambda < 0$ or $\mu < 0$. \\

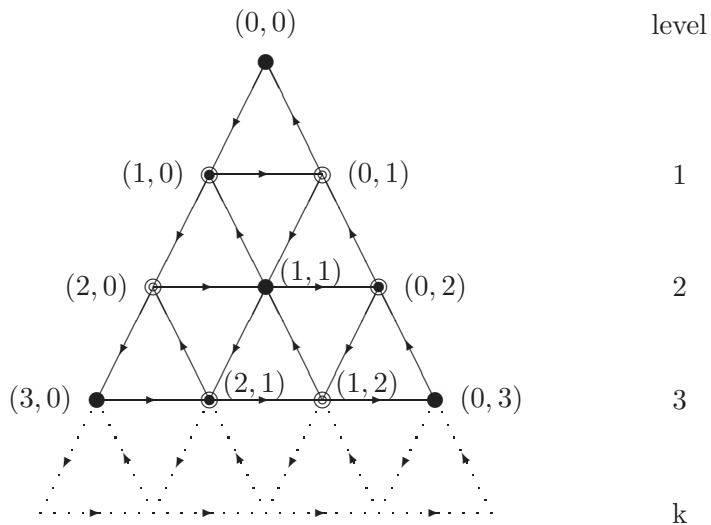
\begin{figure}[h]
\setlength{\unitlength}{1mm}
\begin{center}
\begin{picture}(70,70)

\put(5,0){\begin{picture}(60,60)


\put(0,0){\begin{picture}(15,15)
\dottedline{2}(0,0)(15,0)(7.5,15)(0,0)
\put(7.5,0){\vector(1,0){0.5}}
\put(11.5,7){\vector(-1,2){0,5}}
\put(3.5,7){\vector(-1,-2){0,5}}
\end{picture}}

\put(15,0){\begin{picture}(15,15)
\dottedline{2}(0,0)(15,0)(7.5,15)(0,0)
\put(7.5,0){\vector(1,0){0.5}}
\put(11.5,7){\vector(-1,2){0,5}}
\put(3.5,7){\vector(-1,-2){0,5}}
\end{picture}}

\put(30,0){\begin{picture}(15,15)
\dottedline{2}(0,0)(15,0)(7.5,15)(0,0)
\put(7.5,0){\vector(1,0){0.5}}
\put(11.5,7){\vector(-1,2){0,5}}
\put(3.5,7){\vector(-1,-2){0,5}}
\end{picture}}

\put(45,0){\begin{picture}(15,15)
\dottedline{2}(0,0)(15,0)(7.5,15)(0,0)
\put(7.5,0){\vector(1,0){0.5}}
\put(11.5,7){\vector(-1,2){0,5}}
\put(3.5,7){\vector(-1,-2){0,5}}
\end{picture}}


\put(7.5,15){\begin{picture}(15,15)
\put(0,0){\line(1,0){15}}  
\put(0,0){\line(1,2){7.5}}  
\put(15,0){\line(-1,2){7.5}}
\put(7.5,0){\vector(1,0){0.5}}
\put(11.5,7){\vector(-1,2){0,5}}
\put(3.5,7){\vector(-1,-2){0,5}}
\end{picture}}

\put(22.5,15){\begin{picture}(15,15)
\put(0,0){\line(1,0){15}}  
\put(0,0){\line(1,2){7.5}}  
\put(15,0){\line(-1,2){7.5}}
\put(7.5,0){\vector(1,0){0.5}}
\put(11.5,7){\vector(-1,2){0,5}}
\put(3.5,7){\vector(-1,-2){0,5}}
\end{picture}}

\put(37.5,15){\begin{picture}(15,15)
\put(0,0){\line(1,0){15}}  
\put(0,0){\line(1,2){7.5}}  
\put(15,0){\line(-1,2){7.5}}
\put(7.5,0){\vector(1,0){0.5}}
\put(11.5,7){\vector(-1,2){0,5}}
\put(3.5,7){\vector(-1,-2){0,5}}
\end{picture}}


\put(15,30){\begin{picture}(15,15)
\put(0,0){\line(1,0){15}}  
\put(0,0){\line(1,2){7.5}}  
\put(15,0){\line(-1,2){7.5}}
\put(7.5,0){\vector(1,0){0.5}}
\put(11.5,7){\vector(-1,2){0,5}}
\put(3.5,7){\vector(-1,-2){0,5}}
\end{picture}}

\put(30,30){\begin{picture}(15,15)
\put(0,0){\line(1,0){15}}  
\put(0,0){\line(1,2){7.5}}  
\put(15,0){\line(-1,2){7.5}}
\put(7.5,0){\vector(1,0){0.5}}
\put(11.5,7){\vector(-1,2){0,5}}
\put(3.5,7){\vector(-1,-2){0,5}}
\end{picture}}


\put(22.5,45){\begin{picture}(15,15)
\put(0,0){\line(1,0){15}}  
\put(0,0){\line(1,2){7.5}}  
\put(15,0){\line(-1,2){7.5}}
\put(7.5,0){\vector(1,0){0.5}}
\put(11.5,7){\vector(-1,2){0,5}}
\put(3.5,7){\vector(-1,-2){0,5}}
\end{picture}}


\put(30,60){\circle*{2}}
\put(30,30){\circle*{2}}
\put(7.5,15){\circle*{2}}
\put(52.5,15){\circle*{2}}

\put(22.5,45){\circle*{1.5}}
\put(22.5,45){\circle{2}}
\put(45,30){\circle*{1.5}}
\put(45,30){\circle{2}}
\put(22.5,15){\circle*{1.5}}
\put(22.5,15){\circle{2}}

\put(37.5,45){\circle{2}}
\put(37.5,45){\circle{1}}
\put(15,30){\circle{2}}
\put(15,30){\circle{1}}
\put(37.5,15){\circle{2}}
\put(37.5,15){\circle{1}}

\put(30,65){\makebox(0,0){$(0,0)$}}  
\put(15,45){\makebox(0,0){$(1,0)$}}  
\put(45,45){\makebox(0,0){$(0,1)$}}  
\put(7.5,30){\makebox(0,0){$(2,0)$}}  
\put(52.5,30){\makebox(0,0){$(0,2)$}}  
\put(36,32){\makebox(0,0){$(1,1)$}}  
\put(0,15){\makebox(0,0){$(3,0)$}}  
\put(60,15){\makebox(0,0){$(0,3)$}}  
\put(28.5,17){\makebox(0,0){$(2,1)$}}  
\put(43.5,17){\makebox(0,0){$(1,2)$}}  

\end{picture}}
 
\put(90,65){\makebox(0,0){level}}  
\put(90,45){\makebox(0,0){1}}  
\put(90,30){\makebox(0,0){2}}  
\put(90,15){\makebox(0,0){3}}  
\put(90,0){\makebox(0,0){k}}

\end{picture}
\end{center}
\caption{The $\mathcal{A}_{k}$ graph of $\widehat{su}(3)_k$ for the irrep $(1,0)$.}
\label{graphAk}
\end{figure}

Coefficients of the fusion matrices being non-negative integers, we can naturally consider them as the adjacency matrices of graphs. The correspondance is as follows. The graph associated to $N_{f}$ has $|\mathcal{I}|$ vertices labelled by irreps $\lambda$, and there is $\ell$ 
arcs joigning $\lambda$ to $\mu$ whenever $(N_f)_{\lambda\mu}=\ell$. Since only the fusion matrix associated to the fundamental irrep $(1,0)$ (or its conjugate) is needeed to determine the others, we can focus on the graph related to $N_{(1,0)}$. It is called the $\mathcal{A}_k$
graph for $\widehat{su}(3)_k$ -- the truncated Weyl alcove of $su(3)$ at level $k$ -- and is displayed in Figure \ref{graphAk}. Notice that the graph associated to $(0,1)$ is the same graph with reversed edges.

Triality is well defined on the graph, in the sense that there are only arrows connecting vertices of increasing triality (mod 3). 
Eigenvalues of the adjacency matrix are given by the following
expression:
\begin{equation} 
\gamma^{( \lambda_{1},\lambda_{2}) }=\frac{ 1+ \exp \left(
\frac{2i\pi (\lambda_{1}+1)}{\kappa}\right) + \exp \left( \frac{2i \pi(
\lambda_{1}+1 +\lambda_{2}+1) }{ \kappa } \right)  }{\exp \left(
\frac{2i\pi ( 2(\lambda_{1}+1)+\lambda_{2}+1 ) }{3\kappa }\right)}
\; ,
\qquad \qquad ( \lambda_{1},\lambda_{2}) \in \mathcal{P}_+^k \;.
\label{vp}
\end{equation}
The norm $\beta $ is defined as the maximum eigenvalue. It satisfies 
$2 \leq \beta <3$ (for $k >2$) and its value is given by:
\begin{equation}\label{nrm}
\beta =\gamma^{(0,0) }=\left[ 3 \right]_{q} = 1 + 2 \cos \left( \frac{2\pi}{\kappa}\right) \;,
\end{equation}
where $q=\exp ( i\pi /\kappa) $ is a root of unity and $\left[
x\right] _{q}=\frac{q^{x}-q^{-x}}{q-q^{-1}}$ are q-numbers. 
Notice that in the quantum algebra approach, vertices of the $\mathcal{A}_k$ graphs are 
labelled by irreps of $SU(3)_q$, a finite dimensional quotient of the quantum group 
$U_q(sl(3))$ at roots of unity $q^{2\kappa}=1$. The components of the eigenvector corresponding to $\beta$ (Perron Frobenius vector) give the quantum dimensions of the irreps, and the fusion algebra corresponds to the tensorisation of irreps.


\subsubsection{Fused matrices $F_{\lambda}$ and the Di Francesco-Zuber graphs}

The modular invariants have been classified for the affine $su(2)$ and $su(3)$ models.
The first one bears the name of ADE classification 
\cite{CIZ-class}. This terminology arose from the fact that spectral properties of the $ADE$ Dynkin diagrams (eigenvalues of the adjacency matrix) are encoded in the diagonal terms of the modular invariant $\mathcal{M}$. 
In the same spirit, a first attempt to relate graphs to affine $su(3)$ modular invariants was made in \cite{DiF_Zub-graphs}. The list of proposed graphs was found empirically by ``computed aided flair'' and bears the name of generalized Coxeter graphs or Di Francesco - Zuber diagrams (we refer to \cite{Pet_Zub-CFT} for a list of general conditions that these graphs have to satisfy). The list of graphs proposed in \cite{DiF_Zub-graphs} has later been corrected\footnote{One of these graphs had to be removed because it 
does not satisfy local conditions of cohomological nature.} by Ocneanu, from the condition that these graphs have tu support   a Weak Hopf Algebra structure \cite{Oc-paths}. We refer to \cite{Oc-Bariloche} to the final list.\\

Let $G$ be a graph of this list. The norm of $G$ (maximum eigenvalue of its adjacency matrix)
is:
\begin{equation}
\beta = 1 + 2 \cos \left( \frac{2\pi}{\kappa}\right) \;.
\end{equation}  
This equation defines the generalized Coxeter number $\kappa$ of the graph and its level 
$k=\kappa -3$. The eigenvalues of the graph $G$ are a subset of the eigenvalues of the 
$\mathcal{A}_k$ graph of same level. They are of the form $\gamma^{(r_1,r_2)}$, with $(r_1,r_2) \in Exp(G) \subset \mathcal{P}_+^k$. This set matches with the diagonal terms of the modular invariant associated to the graph. \\ 
  
Vertices of these graphs are labelled by elements of an (integral) basis of irreps of the modules of $SU(3)_q$. 
The vector space $\mathcal{V}(G)$ spanned by the vertices of a graph $G$ is a module under the action of the fusion algebra of the $\mathcal{A}_k$ graph with same level. 
If we note $\lambda, \mu$ the vertices of $\mathcal{A}_k$ and $a,b$ those
of the $G$ graph, we have:
\begin{equation}
\begin{array}{rcl}
\mathcal{A}_k \times \mathcal{V}(G) &\hookrightarrow & \mathcal{V}(G) \\
\lambda \cdot \, a &=& \displaystyle\sum_b (F_{\lambda})_{ab} \; b \;, 
\end{array}
\label{modprop}
\end{equation}
where $(F_{\lambda})_{ab}$ are non-negative integers. The module property 
$(\lambda \cdot \mu) \cdot a = \lambda \cdot (\mu \cdot a)$ implies that the $r \times r$ matrices 
$F_{\lambda}$ -- called fused matrices -- form a representation of dimension $r^2$ of the fusion algebra:
\begin{equation}
F_{\lambda} \, F_{\mu} = \sum_{\nu \in \mathcal{P}_+^k} \mathcal{N}_{\lambda\mu}^{\nu} \, 
F_{\nu} \;.
\label{Fused}
\end{equation}
The fused matrix $F_{(1,0)}$ associated to the fundamental irrep $(1,0)$ is the adjacency matrix 
of the graph $G$. In the CFT approach, the fused matrices give a solution to the Cardy equation \cite{cardy-eq}. In this context, vertices of the graph $G$ can be seen as labelling the boundary conditions of the corresponding CFT (see discussion in \cite{Zuber-CFT,behrend,behrend-BCFT}).   \\

Among the Di Francesco-Zuber graphs, some are not only modules under the graph algebra of the corresponding $\mathcal{A}_k$ graph, but also possess an associative and commutative algebra structure (fusion graph algebra). They are proper subgroups of the corresponding quantum group, in the sense that we can multiply irreps (vertices) of these graphs together and decompose the result into a sum of vertices of the graph. For such a graph with $r$ vertices, we define the $r \times r$ matrices $G_r$. We have a unit $G_0=\munite$ and two generators
$G_1=F_{(1,0)}$ and $G_{1'}=F_{(0,1)}$. Once the multiplication by these generators are known, we can reconstruct the whole table of multiplication of the graph algebra of the graph $G$: 
\begin{equation}
G_a \, G_b = \sum_{c=0}^{r-1} (G_a)_{bc} \, G_c \;,
\end{equation}
with $(G_a)_{bc}$ non-negative integers. These graph algebras are called Pasquier 
algebras \cite{pasquier-ADE} for the $su(2)$ case. The corresponding graphs are called subgroup graphs. In the $su(2)$ case, subgroup graphs are $A_k$, $D_{2k}$, $E_6$ and $E_8$. In the $su(3)$ case, they are the infinite series $\mathcal{A}_k$ (for all $k$), $\mathcal{D}_{k}$ (for $k = 0 \mod3$), $\mathcal{E}_{5}, \mathcal{E}_{9}$ and $\mathcal{E}_{21}$.

\subsection{Intertwiner $E_0$, branching rules, modular operator $\hat{T}$ and the ambichiral algebra $\mathcal{B}$} 
We take a graph $G$ with $r$ vertices as starting point. Its adjacency matrix is 
$F_{(1,0)}$. The other $r \times r$ fused matrices $F_{\lambda}$ are determined from the truncated $SU(3)$ recursion formulae (\ref{su3recform}). We call $\ell$ the number of vertices of the corresponding $\mathcal{A}_k$ graph with same level. 
The essential matrices $E_a$ are $\ell \times r$ matrices defined by:
\begin{equation}
(E_a)_{\lambda b} = (F_{\lambda})_{ab} \;.
\end{equation}
They are related to the notion of essential paths (introduced by Ocneanu in \cite{Oc-paths}). The element $(E_a)_{\lambda b}$ counts the number of essential paths associated to the irrep $\lambda$, starting from the vertex $a$ and ending at the vertex $b$ (see for example the explicit study of the exceptionnal $E_6$ graph of the $su(2)$ case in \cite{Coq-qtetra}). One of them is special, the essential matrix $E_0$ associated to the trivial irrep $\lambda=0$, called intertwiner. It satisfies $N_{(1,0)}E_0 = E_0 F_{(1,0)}$, thus intertwining the adjacency matrices of the $\mathcal{A}_k$ and $G$ graphs. The other essential matrices can be obtained from $E_0$ and from the graph algebras $G_a$ by:
\begin{equation}
E_a = E_0 \, G_a \;.
\end{equation}
\begin{theo}
Let $G$ be a subgroup graph and $\mathcal{A}_k$ the corresponding graph of the $\mathcal{A}$ series of same level. The graph algebra of $G$ is a module under the graph algebra of 
$\mathcal{A}_k$. For $a,b \in G$ and $\lambda \in \mathcal{A}_k$, the branching rules 
$\mathcal{A}_k \hookrightarrow G$ are given by:
\begin{equation}
\lambda \hookrightarrow \sum_b (E_0)_{\lambda b} \, b \;.
\end{equation}
\end{theo}
{\bf Proof:} We write the action of $\lambda \in \mathcal{A}_k$ on $a \in G$:
\begin{eqnarray*}
\lambda \, \cdot \, a = \sum_c (F_{\lambda})_{ac} \, c = \sum_c (E_a)_{\lambda c} \, c
&=& \sum_c (E_0 \, G_a) _{\lambda c} \, c \\
&=& \sum_c \sum_b (E_0)_{\lambda b} \, (G_a)_{bc} \, c
= \sum_b (E_0)_{\lambda b} \; b \, \cdot \, a \;. 
\qquad  \qquad \blacksquare
\end{eqnarray*}

The essential matrix $E_0$ is a rectangular matrix with $(k+1)(k+2)/2$ lines labelled by vertices of $\mathcal{A}_k$ and $r$ columns labelled by vertices of $G$. The restriction 
$\mathcal{A}_k \hookrightarrow G$ (branching rules) of an irrep $\lambda$ is given by the elements of the line of $E_0$ corresponding to $\lambda$. The induction mechanism 
$\mathcal{A}_k \hookleftarrow G$, i.e. the elements $\lambda \in \mathcal{A}_k$ for 
which $a \in G$ appears in the branching rules, is given by elements of the column of $E_0$ corresponding to $a$. The vertices of $\mathcal{A}_k$  have a well defined modular operator value, given by Equation (\ref{Tmodop}). 
We want to assign a modular operator value to the vertices $a$ of the $G$ graph. Suppose $a$ appears in the branching rules of $\lambda$ and $\beta$ of $\mathcal{A}_k$, we can define 
$\hat{T}(a)$ by $\hat{T}(\lambda)$ or $\hat{T}(\beta)$, but this value is ill-defined if
$\hat{T}(\lambda) \not= \hat{T}(\beta)$. We call $\mathcal{B}$ the subset of vertices of $G$ for which the modular operator is well-defined \cite{Coq_Gil-Tmod}. We find that $\mathcal{B}$ is always a sub-algebra of the graph algebra of $G$, and its complement $P$ in the graph is $\mathcal{B}-$ invariant, i.e.:
\begin{equation}
G = \mathcal{B} \oplus P \;, \qquad \qquad \mathcal{B} \cdot \mathcal{B} \subset \mathcal{B} \;, \qquad \qquad
\mathcal{B} \cdot P = P \cdot \mathcal{B} \subset P \;.
\end{equation} 
The algebra $\mathcal{B}$ is called the ambichiral algebra. Using the branching rules $\lambda \hookrightarrow \sum_b (E_0)_{\lambda b} \, b$, we define the extended characters 
$\hat{\chi}_c$, for $c \in \mathcal{B}$, as the following linear combination of affine characters 
$\chi_{\lambda}$:
\begin{equation}
\hat{\chi}_c^k = \sum_{\lambda \in \mathcal{P}_+^k} (E_0)_{\lambda c} \chi_{\lambda}^k \;.
\label{extchar}
\end{equation}

\subsection{Modular invariant $\mathcal{M}$}
The modular invariant partition functions fall in two classes, the block-diagonal (Type I)
and the non block-diagonal (Type II). They are respectively associated to 
subgroup and module graphs. 
For subgroup graphs, the modular invariant is given by:
\begin{equation}
\mathcal{M}_{\lambda\mu} = \sum_{c \in \mathcal{B}} (F_{\lambda})_{0c} \, (F_{\mu})_{0c}
= \sum_{c \in \mathcal{B}} (E_{0})_{\lambda c} (E_{0})_{\mu c} \;, 
\label{typeI-M}
\end{equation}
where $0$ is the unit vertex of the graph $G$ and $\mathcal{B}$ is the subset of the vertices of 
the graph $G$ with well-defined modular operator $\hat{T}$. Defining $(E_{0})^{red}$ as a rectangular $\ell \times r$ matrix, called reduced essential matrix (or reduced intertwiner), deduced from $E_{0}$ by setting to zero all the matrix elements of the columns labelled by vertices not belonging to the subalgebra $\mathcal{B}$, the modular invariant can be put in the simple form $\mathcal{M} = E_0 \, (E_0^{red})^{tr}$. 
The corresponding modular invariant partition function $\mathcal{Z}=\sum_{\lambda,\mu}\chi_{\lambda}\mathcal{M}_{\lambda\mu}\overline{\chi}_{\mu}$ is block diagonal with respect to the extended characters (\ref{extchar}):
\begin{equation}
\mathcal{Z} = \sum_{c \in \mathcal{B}} |\hat{\chi}_c^k|^2 \;.
\label{blockZ}
\end{equation}

To each module graph $G$ corresponds a subgroup graph $H$ called its parent graph. The modular invariant of a module graph is obtained from its parent graph by letting an automorphism $\vartheta$ of vertices of $H$ act on the right components:
\begin{equation}
\mathcal{M}_{\lambda\mu} = \sum_{c \in \mathcal{B}} (F_{\lambda})_{0c} \, (F_{\mu})_{0\vartheta(c)}
= \sum_{c \in \mathcal{B}} (E_{0})_{\lambda c} (E_{0})_{\mu \vartheta(c)} \;. 
\label{typeII-M}
\end{equation}
The involution $\vartheta$ is such that it preserves the value of the modular operator: 
$\hat{T}[\vartheta(c)]=\hat{T}[c]$, for $c \in \mathcal{B}$. 
It leads to non block-diagonal partition functions:
\begin{equation}
\mathcal{Z}  = \sum_{c \in \mathcal{B}} \, \hat{\chi}_c^k \; \overline{\hat{\chi}_{\vartheta(c)}^k} \;.
\label{nonblockZ}
\end{equation}
All modular invariant partition functions are obtained from subgroup graphs and from involutions $\vartheta$. We do not discuss the module graphs and how they are constructed. 
In plain terms, given a subgroup graph, we look for all involutions preserving the modular operator value. This is a necessary but not sufficient condition. It leads to a $\mathcal{M}$ that commutes with the modular $\mathcal{T}$ matrix, but does not garantee that it commutes with the modular $\mathcal{S}$ matrix.  The determination of such involutions is made by an explicit analysis of the values of the modular operator on the subset $\mathcal{B}$ of vertices of the graph $G$, and then by checking the commutativity with the $\mathcal{S}$ matrix.
The suitable involutions $\vartheta$ are the conjugation $C$, a twist $\rho$ related to the $\mathbb{Z}_3$-symmetry or the combination of conjugation and twist. The explicit analysis is made in the next sections.

\subsection{Comments about quantum symmetries and Ocneanu graphs}
In this article, the modular invariants are determined by the formulae
(\ref{typeI-M}) and (\ref{typeII-M}), leading to the block-diagonal (\ref{blockZ})
and the non block-diagonal (\ref{nonblockZ}) partition functions.
The validity of these expressions comes from an analysis of the algebra of quantum symmetries
of the Di Francesco-Zuber graphs, which is not the purpose of this paper. Let us just mention 
the following. Given such a graph $G$, 
the Ocneanu construction associates to it a special kind of finite dimensional weak Hopf algebra (WHA) \cite{Oc-paths} (see also \cite{Pet_Zub-Oc,Gil-tesis,Coq_Trinchero}). This WHA has a product $\circ$ and a product $\hat{\circ}$ on the dual space, defined from the coproduct by a pairing. It is a finite dimensional semi-simple algebra for both structures. The two algebra structures give rise to two algebras of characters. The first one is a commutative and associative algebra, with one generator (plus its conjugated), isomorphic to the fusion algebra of the corresponding affine $su(3)$ CFT, and is encoded by the $\mathcal{A}_k$ graph of same level of $G$. The second one, called Ocneanu algebra of quantum symmetries, is an algebra with two generators and their conjugated (only one if $G=\mathcal{A}$). It is an associative but not always commutative algebra. Multiplication by the generators is encoded by a graph called the Ocneanu graph of $G$. The list of Ocneanu graphs for the $SU(3)$ system is known but was never made available in written form\footnote{Several cases are analyzed in \cite{Coq_Gil-Tmod, Gil-tesis}, but we plan to give the full list in a forthcoming publication \cite{Qsymsu3}.}. 
The algebra of quantum symmetries $Oc(G)$ of a subgroup graph $G$ can be realized as a quotient of the tensor square of the graph algebra of $G$ \cite{Coq-qtetra,Coq_Gil-ADE, Gil-tesis}. Elements $x \in Oc(G)$ are writen
as $a \otimes_\mathcal{B} b$, where $a,b \in G$ and $\mathcal{B}$ is the ambichiral algebra characterized by modular properties. The algebra $Oc(G)$ has a bimodule structure under the left-right action of the graph algebra of the corresponding $\mathcal{A}_k$. For $\lambda,\mu \in \mathcal{A}_k$, we have:
\begin{equation}
\lambda \cdot x \cdot \mu = \sum_y \mathcal{W}_{\lambda\mu \,, xy} \; y \;,
\label{dfcoef}
\end{equation}
where $\mathcal{W}_{\lambda\mu \,, xy}$ are called the double fusion coefficients. The double fusion matrices $V_{\lambda\mu}$ are defined by $(V_{\lambda\mu})_{xy} = 
\mathcal{W}_{\lambda\mu \,, xy}$. 
The associativity property: $ \lambda \cdot (\lambda' \cdot x \cdot \mu) \cdot \mu' = 
(\lambda \cdot \lambda') \cdot x \cdot  (\mu \cdot \mu')$ of the bimodule structure implies that  the double fusion matrices satisfy:
\begin{equation}
V_{\lambda \mu} \; V_{\lambda' \mu'} = \sum_{\lambda'', \mu'' \in \mathcal{P}_+^{r,k}} 
\; \mathcal{N}_{\lambda\lambda'}^{\lambda''} \;  
\mathcal{N}_{\mu\mu'}^{\mu''} \; V_{\lambda''\mu''}. 
\label{modsplit}
\end{equation}
This is exactly the compatibility relation that have to satisfy the generalized partition functions, which are defined by $\mathcal{Z}_{xy} = \sum_{\lambda,\mu}
\chi_{\lambda}\mathcal{W}_{\lambda\mu \,, xy}\overline{\chi_{\mu}}$. They correspond to systems with two defect lines labelled by $x$ and $y$. The modular invariant corresponds to the case with no defect lines $x=y=0$. We can show that Equation (\ref{dfcoef}) leads to the expression (\ref{typeI-M}) used in this paper \cite{Gil-tesis}. The algebra of quantum symmetries of module graphs are constructed from a twist of the algebra of quantum symmetries of a subgroup graph, called its parent graph (see for example \cite{Pet_Zub-Oc,Coq_Gil-Tmod, Gil-tesis}). A similar discussion can be made for the corresponding Type II modular invariants, leading to the expression (\ref{typeII-M}). For example, the complete $su(2)$ model is analyzed in this way in \cite{Gil-tesis}. Finally, let us mention that Equation (\ref{modsplit}) can actually be used as a starting point if we are given the list of modular invariants $\mathcal{M}$ (see an example in \cite{Coq_Esteban} and a forthcoming publication \cite{Gil_Esteban}). All Ocneanu graphs can be obtained in this way, and the subgroup and module graphs can be deduced from their Ocneanu graphs.


\section{The $\mathcal{A}_k$ graphs}

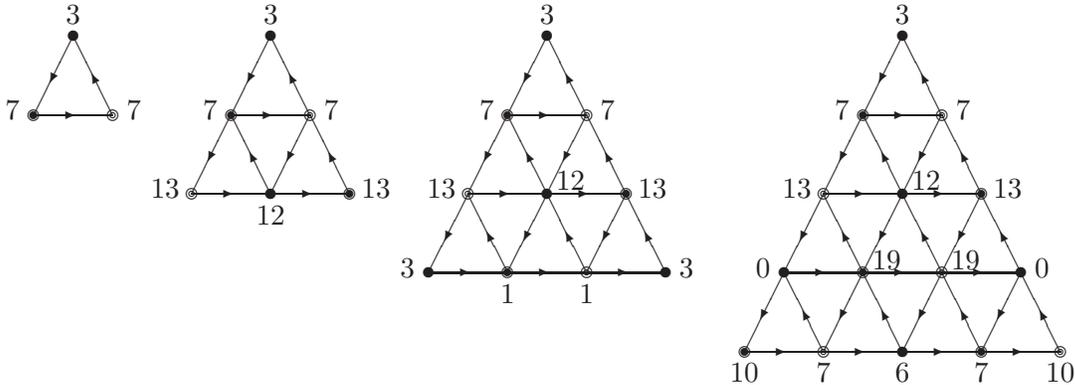
\begin{figure}[h]
\setlength{\unitlength}{0.7mm}
\begin{center}
\begin{picture}(195,70)
\put(0,45){\begin{picture}(15,15)
\put(0,0){\line(1,0){15}}  
\put(0,0){\line(1,2){7.5}}  
\put(15,0){\line(-1,2){7.5}}
\put(7.5,0){\vector(1,0){0.5}}
\put(11.5,7){\vector(-1,2){0,5}}
\put(3.5,7){\vector(-1,-2){0,5}}
\put(7.5,15){\circle*{2}}
\put(0,0){\circle*{1.5}}
\put(0,0){\circle{2}}
\put(15,0){\circle{2}}
\put(15,0){\circle{1}}
\end{picture}}

\put(30,30){\begin{picture}(30,30)

\put(0,0){\begin{picture}(15,15)
\put(0,0){\line(1,0){15}}  
\put(0,0){\line(1,2){7.5}}  
\put(15,0){\line(-1,2){7.5}}
\put(7.5,0){\vector(1,0){0.5}}
\put(11.5,7){\vector(-1,2){0,5}}
\put(3.5,7){\vector(-1,-2){0,5}}
\end{picture}}

\put(15,0){\begin{picture}(15,15)
\put(0,0){\line(1,0){15}}  
\put(0,0){\line(1,2){7.5}}  
\put(15,0){\line(-1,2){7.5}}
\put(7.5,0){\vector(1,0){0.5}}
\put(11.5,7){\vector(-1,2){0,5}}
\put(3.5,7){\vector(-1,-2){0,5}}
\end{picture}}

\put(7.5,15){\begin{picture}(15,15)
\put(0,0){\line(1,0){15}}  
\put(0,0){\line(1,2){7.5}}  
\put(15,0){\line(-1,2){7.5}}
\put(7.5,0){\vector(1,0){0.5}}
\put(11.5,7){\vector(-1,2){0,5}}
\put(3.5,7){\vector(-1,-2){0,5}}
\end{picture}}

\put(15,30){\circle*{2}}
\put(7.5,15){\circle*{1.5}}
\put(7.5,15){\circle{2}}
\put(22.5,15){\circle{2}}
\put(22.5,15){\circle{1}}
\put(0,0){\circle{2}}
\put(0,0){\circle{1}}
\put(30,0){\circle*{1.5}}
\put(30,0){\circle{2}}
\put(15,0){\circle*{2}}

\end{picture}}

\put(75,15){\begin{picture}(45,45)

\put(0,0){\begin{picture}(15,15)
\put(0,0){\line(1,0){15}}  
\put(0,0){\line(1,2){7.5}}  
\put(15,0){\line(-1,2){7.5}}
\put(7.5,0){\vector(1,0){0.5}}
\put(11.5,7){\vector(-1,2){0,5}}
\put(3.5,7){\vector(-1,-2){0,5}}
\end{picture}}

\put(15,0){\begin{picture}(15,15)
\put(0,0){\line(1,0){15}}  
\put(0,0){\line(1,2){7.5}}  
\put(15,0){\line(-1,2){7.5}}
\put(7.5,0){\vector(1,0){0.5}}
\put(11.5,7){\vector(-1,2){0,5}}
\put(3.5,7){\vector(-1,-2){0,5}}
\end{picture}}

\put(30,0){\begin{picture}(15,15)
\put(0,0){\line(1,0){15}}  
\put(0,0){\line(1,2){7.5}}  
\put(15,0){\line(-1,2){7.5}}
\put(7.5,0){\vector(1,0){0.5}}
\put(11.5,7){\vector(-1,2){0,5}}
\put(3.5,7){\vector(-1,-2){0,5}}
\end{picture}}

\put(7.5,15){\begin{picture}(15,15)
\put(0,0){\line(1,0){15}}  
\put(0,0){\line(1,2){7.5}}  
\put(15,0){\line(-1,2){7.5}}
\put(7.5,0){\vector(1,0){0.5}}
\put(11.5,7){\vector(-1,2){0,5}}
\put(3.5,7){\vector(-1,-2){0,5}}
\end{picture}}

\put(22.5,15){\begin{picture}(15,15)
\put(0,0){\line(1,0){15}}  
\put(0,0){\line(1,2){7.5}}  
\put(15,0){\line(-1,2){7.5}}
\put(7.5,0){\vector(1,0){0.5}}
\put(11.5,7){\vector(-1,2){0,5}}
\put(3.5,7){\vector(-1,-2){0,5}}
\end{picture}}

\put(15,30){\begin{picture}(15,15)
\put(0,0){\line(1,0){15}}  
\put(0,0){\line(1,2){7.5}}  
\put(15,0){\line(-1,2){7.5}}
\put(7.5,0){\vector(1,0){0.5}}
\put(11.5,7){\vector(-1,2){0,5}}
\put(3.5,7){\vector(-1,-2){0,5}}
\end{picture}}

\put(0,0){\circle*{2}}
\put(45,0){\circle*{2}}
\put(22.5,15){\circle*{2}}
\put(22.5,45){\circle*{2}}

\put(15,0){\circle*{1.5}}
\put(15,0){\circle{2}}
\put(15,30){\circle*{1.5}}
\put(15,30){\circle{2}}
\put(37.5,15){\circle*{1.5}}
\put(37.5,15){\circle{2}}

\put(30,0){\circle{2}}
\put(30,0){\circle{1}}
\put(7.5,15){\circle{2}}
\put(7.5,15){\circle{1}}
\put(30,30){\circle{2}}
\put(30,30){\circle{1}}

\end{picture}}

\put(135,0){\begin{picture}(60,60)

\put(0,0){\begin{picture}(15,15)
\put(0,0){\line(1,0){15}}  
\put(0,0){\line(1,2){7.5}}  
\put(15,0){\line(-1,2){7.5}}
\put(7.5,0){\vector(1,0){0.5}}
\put(11.5,7){\vector(-1,2){0,5}}
\put(3.5,7){\vector(-1,-2){0,5}}
\end{picture}}

\put(15,0){\begin{picture}(15,15)
\put(0,0){\line(1,0){15}}  
\put(0,0){\line(1,2){7.5}}  
\put(15,0){\line(-1,2){7.5}}
\put(7.5,0){\vector(1,0){0.5}}
\put(11.5,7){\vector(-1,2){0,5}}
\put(3.5,7){\vector(-1,-2){0,5}}
\end{picture}}

\put(30,0){\begin{picture}(15,15)
\put(0,0){\line(1,0){15}}  
\put(0,0){\line(1,2){7.5}}  
\put(15,0){\line(-1,2){7.5}}
\put(7.5,0){\vector(1,0){0.5}}
\put(11.5,7){\vector(-1,2){0,5}}
\put(3.5,7){\vector(-1,-2){0,5}}
\end{picture}}

\put(45,0){\begin{picture}(15,15)
\put(0,0){\line(1,0){15}}  
\put(0,0){\line(1,2){7.5}}  
\put(15,0){\line(-1,2){7.5}}
\put(7.5,0){\vector(1,0){0.5}}
\put(11.5,7){\vector(-1,2){0,5}}
\put(3.5,7){\vector(-1,-2){0,5}}
\end{picture}}

\put(7.5,15){\begin{picture}(15,15)
\put(0,0){\line(1,0){15}}  
\put(0,0){\line(1,2){7.5}}  
\put(15,0){\line(-1,2){7.5}}
\put(7.5,0){\vector(1,0){0.5}}
\put(11.5,7){\vector(-1,2){0,5}}
\put(3.5,7){\vector(-1,-2){0,5}}
\end{picture}}

\put(22.5,15){\begin{picture}(15,15)
\put(0,0){\line(1,0){15}}  
\put(0,0){\line(1,2){7.5}}  
\put(15,0){\line(-1,2){7.5}}
\put(7.5,0){\vector(1,0){0.5}}
\put(11.5,7){\vector(-1,2){0,5}}
\put(3.5,7){\vector(-1,-2){0,5}}
\end{picture}}

\put(37.5,15){\begin{picture}(15,15)
\put(0,0){\line(1,0){15}}  
\put(0,0){\line(1,2){7.5}}  
\put(15,0){\line(-1,2){7.5}}
\put(7.5,0){\vector(1,0){0.5}}
\put(11.5,7){\vector(-1,2){0,5}}
\put(3.5,7){\vector(-1,-2){0,5}}
\end{picture}}

\put(15,30){\begin{picture}(15,15)
\put(0,0){\line(1,0){15}}  
\put(0,0){\line(1,2){7.5}}  
\put(15,0){\line(-1,2){7.5}}
\put(7.5,0){\vector(1,0){0.5}}
\put(11.5,7){\vector(-1,2){0,5}}
\put(3.5,7){\vector(-1,-2){0,5}}
\end{picture}}

\put(30,30){\begin{picture}(15,15)
\put(0,0){\line(1,0){15}}  
\put(0,0){\line(1,2){7.5}}  
\put(15,0){\line(-1,2){7.5}}
\put(7.5,0){\vector(1,0){0.5}}
\put(11.5,7){\vector(-1,2){0,5}}
\put(3.5,7){\vector(-1,-2){0,5}}
\end{picture}}

\put(22.5,45){\begin{picture}(15,15)
\put(0,0){\line(1,0){15}}  
\put(0,0){\line(1,2){7.5}}  
\put(15,0){\line(-1,2){7.5}}
\put(7.5,0){\vector(1,0){0.5}}
\put(11.5,7){\vector(-1,2){0,5}}
\put(3.5,7){\vector(-1,-2){0,5}}
\end{picture}}

\put(30,0){\circle*{2}}
\put(30,30){\circle*{2}}
\put(30,60){\circle*{2}}
\put(7.5,15){\circle*{2}}
\put(52.5,15){\circle*{2}}

\put(0,0){\circle*{1.5}}
\put(0,0){\circle{2}}
\put(45,0){\circle*{1.5}}
\put(45,0){\circle{2}}
\put(22.5,15){\circle*{1.5}}
\put(22.5,15){\circle{2}}
\put(45,30){\circle*{1.5}}
\put(45,30){\circle{2}}
\put(22.5,45){\circle*{1.5}}
\put(22.5,45){\circle{2}}

\put(15,0){\circle{2}}
\put(15,0){\circle{1}}
\put(60,0){\circle{2}}
\put(60,0){\circle{1}}
\put(15,30){\circle{2}}
\put(15,30){\circle{1}}
\put(37.5,15){\circle{2}}
\put(37.5,15){\circle{1}}
\put(37.5,45){\circle{2}}
\put(37.5,45){\circle{1}}

\end{picture}}


\put(7.5,64){\makebox(0,0){3}}
\put(45,64){\makebox(0,0){3}}
\put(97.5,64){\makebox(0,0){3}}
\put(165,64){\makebox(0,0){3}}

\put(-4,46){\makebox(0,0){7}}
\put(19,46){\makebox(0,0){7}}
\put(33.5,46){\makebox(0,0){7}}
\put(56.5,46){\makebox(0,0){7}}
\put(86,46){\makebox(0,0){7}}
\put(109,46){\makebox(0,0){7}}
\put(153.5,46){\makebox(0,0){7}}
\put(176.5,46){\makebox(0,0){7}}

\put(25,31){\makebox(0,0){13}}
\put(65,31){\makebox(0,0){13}}
\put(77.5,31){\makebox(0,0){13}}
\put(117.5,31){\makebox(0,0){13}}
\put(145,31){\makebox(0,0){13}}
\put(185,31){\makebox(0,0){13}}
\put(45,26){\makebox(0,0){12}}
\put(102,32.5){\makebox(0,0){12}}
\put(169.5,32.5){\makebox(0,0){12}}

\put(71,16){\makebox(0,0){3}}
\put(124,16){\makebox(0,0){3}}
\put(138.5,16){\makebox(0,0){0}}
\put(191.5,16){\makebox(0,0){0}}
\put(90,11){\makebox(0,0){1}}
\put(105,11){\makebox(0,0){1}}
\put(162,17.5){\makebox(0,0){19}}
\put(177,17.5){\makebox(0,0){19}}

\put(135,-4){\makebox(0,0){10}}
\put(150,-4){\makebox(0,0){7}}
\put(165,-4){\makebox(0,0){6}}
\put(180,-4){\makebox(0,0){7}}
\put(195,-4){\makebox(0,0){10}}

\end{picture}


\end{center}
\caption{The $\mathcal{A}_{k}$ graphs for k=1 to 4 and the values of the modular operator on their vertices. The labelling is such that the unit vertex (0,0) is on the top of the triangle.}
\label{fig:ak1234}
\end{figure}

The $\mathcal{A}_k$ graphs have $(k+1)(k+2)/2$ vertices labelled by integral highest weight representations of $\widehat{su}(3)_k$. We recall that each vertex has a well-defined modular operator value, given by:
\begin{equation}
\hat{T}[(\lambda_1,\lambda_2)] = (\lambda_1+1)^2 + (\lambda_1+1)(\lambda_2+1) + 
(\lambda_2+1)^2 \qquad \qquad \mod 3 \kappa \;.
\end{equation} 
The $\mathcal{A}_k$ graphs are displayed in Figure \ref{fig:ak1234} for $k=1$ to 4,  together with the values of $\hat{T}$ on their vertices.

The adjacency matrix of the $\mathcal{A}_k$ graph is $N_{(1,0)}$. The $\mathcal{A}_k$ graph is a trivial module under itself, the fused matrices $F_{\lambda}$ are equal to the fusion matrices $N_{\lambda}$. Branching rules $\mathcal{A}_k \hookrightarrow
\mathcal{A}_k$ are trivial, the essential matrix $E_{(0,0)}$ is the unit matrix. All vertices of 
$\mathcal{A}_k$ having well-defined modular operator value, we set $\mathcal{B}=\mathcal{A}_k$.
The modular invariant is diagonal 
$\mathcal{M}_{\lambda\mu} = \delta_{\lambda \mu}$
and the modular invariant partition function associated to the $\mathcal{A}_k$ graph is:
\begin{equation}
\mathcal{Z}(\mathcal{A}_k) = \sum_{\lambda \in \mathcal{P}_+^k} |\chi_{\lambda}^k|^2 \;,
\qquad \qquad \qquad \qquad \qquad \forall k\geq 1 \;.
\end{equation}

\paragraph{The conjugated $\mathcal{A}_k$ series}

The value of the modular operator $\hat{T}$ on conjugated vertices of $\mathcal{A}_k$ is the same: $\hat{T}[\lambda] = \hat{T}[\lambda^*]$. The conjugation $C$ is a $\hat{T}$-invariant involution. In Figure \ref{fig:ak1234}, it corresponds to the reflexion by the vertical axis. We get the modular invariant $\mathcal{M}_{\lambda\mu} = \delta_{\lambda \mu^*}$, corresponding to the conjugated graph $\mathcal{A}_k^*$, and the corresponding modular invariant partition function:
\begin{equation}
\mathcal{Z}(\mathcal{A}_k^*) = \sum_{\lambda \in \mathcal{P}_+^k} \, \chi_{\lambda}^k \; 
\overline{\chi_{\lambda^*}^{k}}\;,
\qquad \qquad \qquad \qquad \qquad \forall k\geq 1 \;.
\end{equation}
Notice that we do not discuss or even give the conjugated graph $\mathcal{A}_k^*$ (see for example \cite{Evans-conj} for a discussion of the construction of conjugated graphs.) The corresponding partition function is obtained here from the conjugation involution defined on vertices of the $\mathcal{A}_k$ graph.


\paragraph{The twisted $\mathcal{A}_k$ series}
The twist $\rho$ of the vertices of $\mathcal{A}_k$ is defined by $\rho(\lambda)=
A^{kt(\lambda)}(\lambda)$, where $k$ is the level, $t(\lambda)$ is the triality and $A$ is the $\ZZ_3$ symmetry
of the $\mathcal{A}_k$ graph. The twist $\rho$ is a $\hat{T}$-invariant involution, i.e. 
$\rho^2=\munite$ and $\hat{T}[\rho(\lambda)]=\hat{T}[\lambda]$. 
The modular invariant of the twisted $\mathcal{A}_k$ model is given by  
$\mathcal{M}_{\lambda\mu}  = \delta_{\lambda \rho(\mu)}$.
The corresponding Di Francesco-Zuber graphs are the $\mathbb{Z}_3$-orbifold graphs $\mathcal{D}_k = \mathcal{A}_k/3$. Notice that, for $k=0\mod3$, $\rho(\lambda)=\lambda$, so the twist does not produce a new modular invariant. In fact, the orbifold graphs $\mathcal{D}_k$ are module graphs for $k=1,2$ (mod 3), but they are subgroup graphs for $k=0$ (mod 3). In the subgroup case, the corresponding modular invariant is not obtained from a twist of the $\mathcal{A}_k$ graph. The orbifold procedure and the subgroup graphs are analyzed in the next section.
The modular invariant partition functions corresponding to the orbifold graphs 
$\mathcal{D}_k$, for $k\not= 0$ (mod 3), are given by:
\begin{equation}
\mathcal{Z}(\mathcal{A}_k^{\rho}) = \sum_{\lambda \in \mathcal{P}_+^k} \, \chi_{\lambda}^k \; 
\overline{\chi_{\rho(\lambda)}^{k}}\;,
\qquad \qquad \qquad \qquad \qquad \textrm{for } k \not= 0 \mod 3  \;.
\end{equation}
For $k=1$, we have 3 irreps $(0,0), (1,0)$ and $(0,1)$ and the twisted vertices are:
$\rho(0,0)=(0,0)$, $\rho(1,0)=(0,1)$ and $\rho(0,1)=(1,0)$. So we have 
$\rho(\lambda)=\lambda^*$, and the twisted modular invariant is equal to the conjugated one. 
This is also the case for $k=2$. Conjugate and orbifold graphs are equal for $k=1,2$:
$\mathcal{D}_1=\mathcal{A}_1^*$, $\mathcal{D}_2=\mathcal{A}_2^*$.

\paragraph{The $\mathcal{A}_4$ twist} The twist $\rho$ on the vertices of the 
$\mathcal{A}_4$ graph (see Figure \ref{fig:ak1234}) gives: 
\begin{equation*}
\begin{array}{rclcrclcrcl}
\rho(0,0) &=& (0,0) & \qquad \qquad & \rho(1,0) &=& (3,1) & \qquad \qquad & \rho(3,1) &=& (1,0) \\
\rho(1,1) &=& (1,1) & & \rho(2,1) &=& (1,2) & & \rho(1,2) &=& (2,1) \\
\rho(2,2) &=& (2,2) & & \rho(4,0) &=& (0,4) & & \rho(0,4) &=& (4,0) \\
\rho(3,0) &=& (3,0) & & \rho(0,2) &=& (2,0) & & \rho(2,0) &=& (0,2) \\
\rho(0,3) &=& (0,3) & & \rho(1,3) &=& (0,1) & & \rho(0,1) &=& (1,3) 
\end{array}
\end{equation*}
We can verify that the twisted vertex $\rho(\lambda)$ have the same operator value $\hat{T}$
than the vertex $\lambda$. The modular invariant is determined by $\mathcal{M}_{\lambda\mu}  = \delta_{\lambda \rho(\mu)}$. Choosing the following order for the vertices of $\mathcal{A}_k$:
(0,0), (1,0), (0,1), (2,0), (1,1), (0,2), (3,0), (2,1), (1,2), (0,3), (4,0), (3,1), (2,2),
(1,3), (0,4), we get:
\begin{equation}
\mathcal{M} = \left(
\begin{array}{ccccccccccccccc}
1 & . & . & . & . & . & . & . & . & . & . & . & . & . & . \\
. & . & . & . & . & . & . & . & . & . & . & 1 & . & . & . \\
. & . & . & . & . & . & . & . & . & . & . & . & . & 1 & . \\
. & . & . & . & . & 1 & . & . & . & . & . & . & . & . & . \\
. & . & . & . & 1 & . & . & . & . & . & . & . & . & . & . \\
. & . & . & 1 & . & . & . & . & . & . & . & . & . & . & . \\
. & . & . & . & . & . & 1 & . & . & . & . & . & . & . & . \\
. & . & . & . & . & . & . & . & 1 & . & . & . & . & . & . \\ 
. & . & . & . & . & . & . & 1 & . & . & . & . & . & . & . \\
. & . & . & . & . & . & . & . & . & 1 & . & . & . & . & . \\
. & . & . & . & . & . & . & . & . & . & . & . & . & . & 1 \\
. & 1 & . & . & . & . & . & . & . & . & . & . & . & . & . \\
. & . & . & . & . & . & . & . & . & . & . & . & 1 & . & . \\
. & . & 1 & . & . & . & . & . & . & . & . & . & . & . & . \\
. & . & . & . & . & . & . & . & . & . & 1 & . & . & . & .
\end{array}
\right)
\end{equation}
We can check that this matrix commutes with the modular matrices $\mathcal{S}$ and 
$\mathcal{T}$, the graph associated is the module orbifold graph 
$\mathcal{D}_4=\mathcal{A}_4/3$.


\paragraph{The twisted conjugated $\mathcal{A}_k$ series}
Combining the two $\hat{T}$-invariant involutions $\rho$ and $C$ also gives a 
$\hat{T}$-invariant involution: $\hat{T}[\rho(\lambda)^*] = \hat{T}[\lambda]$. 
The corresponding modular invariant is given by:
$\mathcal{M}_{\lambda\mu} = \delta_{\lambda \rho(\mu)^*}$. This is valid for $k\not=0$ (mod 3).
The conjugation of the subgroup orbifold graphs $\mathcal{D}_k$ for $k=0$ (mod 3) are discused in the next section. 
The modular invariant partition functions of the conjugated twisted $\mathcal{A}_k$ model, for $k\not=0$ (mod 3), associated to the graphs $\mathcal{D}_k^*$, are given by:
\begin{equation}
\mathcal{Z}(\mathcal{D}_k^*) = \sum_{\lambda \in \mathcal{P}_+^k} \chi_{\lambda}^k \; 
\overline{\chi_{\rho(\lambda)^*}^{k}} 
\qquad \qquad \qquad \qquad \qquad \textrm{for } k \not= 0 \mod 3 \;.
\end{equation}
For $k=1$ and 2, we have $\rho(\lambda)^* = \lambda$, the twisted conjugated modular invariant is equal to the diagonal one. Conjugate-orbifold graphs are equal to $\mathcal{A}_k$ for 
$k=1,2$: $\mathcal{D}_1^*=\mathcal{A}_1$, $\mathcal{D}_2^*=\mathcal{A}_2^*$.


\section{The orbifold $\mathcal{D}_k=\mathcal{A}_k/3$ graphs}
The $\mathcal{D}_k$ graphs are the $\mathbb{Z}_3$-orbifold of the $\mathcal{A}_k$ graphs and are also denoted $\mathcal{A}_k/3$. The orbifold procedure is described in \cite{Kostov,Fendley, DiF_Zub-graphs}. For $k\not= 0 \mod3$, these graphs are module graphs and the corresponding modular invariant is obtained from a twist of the $\mathcal{A}_k$ graph of same level (see previous section). For $k=0\mod 3$, they are subgroup graphs, and we denote them 
$\mathcal{D}_{k\equiv 0}$. In the rest of this section, we will only consider the subgroup case, hence take $k=0\mod3$. The $\mathbb{Z}_3$-symmetry is the automorphism $A$ defined in Equation (\ref{z3sym}).
The central vertex $\alpha=(\frac{k}{3},\frac{k}{3})$ of $\mathcal{A}_k$ is $\mathbb{Z}_3$-invariant: $A(\alpha)=\alpha$. The $\mathbb{Z}_3$ symmetry corresponds to a rotation of angle $\pi/3$ around this central vertex $\alpha$. For $\lambda \not= \alpha$, the 3 vertices $\lambda,A(\lambda)$ and $A^2(\lambda)$ belong to the same orbit in $\mathcal{A}_k$ and lead to a single vertex in the orbifold graph $\mathcal{D}_{k\equiv 0}$, denoted by $\tilde{\lambda}$. 
The central vertex $\alpha$ (fixed point) is triplicated\footnote{For $k \not= 0 \mod3$, there is no fixed point under the $\mathbb{Z}_3$-symmetry. The orbifold procedure is more simple in this case.} in the orbifold graph.  We introduce 3 copies $\alpha^{(i)}$, $i=1,2,3$, for the vertex $\alpha$. The number of vertices of $\mathcal{D}_{k\equiv 0}$ is therefore equal to $(\frac{(k+1)(k+2)}{2}-1)/3+3$. Vertices $a \in \mathcal{D}_{k\equiv 0}$ are denoted $\{\tilde{\lambda},\alpha^{(i)}\}$. The adjacency matrix of
$\mathcal{D}_{k\equiv 0}$ is constructed as follows. 
Let $N=N_{(1,0)}$ be the adjacency matrix of the 
$\mathcal{A}_k$ graph. From the $\mathbb{Z}_3$-symmetry of this graph we have 
$N_{\lambda\mu}=N_{A(\lambda)A(\mu)}$.  Consider the following matrix:
\begin{subequations} 
\begin{eqnarray}
G_{\lambda\mu} = \dps \sum_{a=0}^{2} N_{\lambda A^a(\mu)}\;, \qquad \qquad \qquad \qquad \qquad \qquad \qquad \qquad \quad 
&\textrm{ for }& \dps \lambda, \mu \not= \alpha  \\
G_{\lambda\alpha^{(i)}} = N_{\lambda\alpha} \,, \quad G_{\alpha^{(i)}\lambda} = 
N_{\alpha\lambda} \,, \quad G_{\alpha^{(i)}\alpha^{(i)}} = N_{\alpha\alpha}=0 \,, \qquad 
&\textrm{ for }&  \lambda \not= \alpha 
\end{eqnarray}
\label{adjmatrorbifold}
\end{subequations}
It has the following property: $G_{\lambda\mu}=G_{\lambda A(\mu)}=G_{A(\lambda)\mu}$. Lines and columns corresponding to vertices of the same orbit are equal. The adjacency matrix 
of the $\mathcal{D}_{k\equiv 0}$ graph is obtained from $G_{\lambda\mu}$ after identifying each point $\lambda$ with its orbit $\tilde{\lambda}$, i.e. eliminating identical lines and columns: 
$D_{\tilde{\lambda}\tilde{\mu}}=G_{\lambda\mu}$.\\

The $\mathcal{D}_{k\equiv 0}$ are subgroup graphs. We can define a multiplication between vertices of these graphs, with non-negative integer coefficient structures. The adjacency matrix 
encodes multiplication by the left generator, and with these data we can reconstruct the whole table of multiplication. The vector space generated by the vertices of $\mathcal{D}_{k\equiv 0}$ is a module under the graph algebra of $\mathcal{A}_k$. The fused matrices $F_{\lambda}$
coding this module property are obtained by the truncated $SU(3)$ recursion formulae, with
starting point $F_{(1,0)}$ given by the adjacency matrix of $\mathcal{D}_k$. We easily obtain the essential matrix (intertwiner) $E_{\tilde{0}}$. The branching rules $\mathcal{A}_k \hookrightarrow \mathcal{D}_k$, obtained from $E_{\tilde{0}}$, are given by:
$\lambda \hookrightarrow \tilde{\lambda}, \alpha \hookrightarrow \alpha^{(1)}+\alpha^{(2)}+\alpha^{(3)}$. The corresponding induction rules $\mathcal{D}_k \hookleftarrow \mathcal{A}_k$ are:
\begin{equation}
\tilde{\lambda} \hookleftarrow \lbrace \lambda, A(\lambda), A^2(\lambda) \rbrace \;, \qquad \qquad
\alpha^{(i)} \hookleftarrow \alpha \;.
\end{equation}
Notice that vertices belonging to the same orbit have the same triality: $t(A(\lambda_1,\lambda_2)) = t(\lambda_1,\lambda_2) + 
(k-3\lambda_1) = t(\lambda)\mod3$. Triality is therefore well defined on the orbifold graphs
$\mathcal{D}_{k\equiv 0}$. 
The modular operator $\hat{T}$ has the following value on the vertex $A(\lambda)$:
\begin{equation}
\hat{T}[A(\lambda_1,\lambda_2)] = \hat{T}[(k-\lambda_1-\lambda_2,\lambda_1)] = \hat{T}[\lambda]
+(k+3)(k-3\lambda_2-t(\lambda)) \;.
\end{equation}
We have $\hat{T}[A(\lambda)]=\hat{T}(\lambda)$ only for vertices with $t(\lambda)=0$. From the induction mechanism, the subset $\mathcal{B}$ of vertices $a \in \mathcal{D}_{k\equiv 0}$ with well-defined modular operator are those with triality equal to 0:
\begin{equation}
\mathcal{B}=\{ \alpha^{(i)} , \tilde{\lambda} \,|\, t(\tilde{\lambda})=0 \} \;.
\end{equation}
We can check that the multiplication of elements of $\mathcal{B}$ is closed: $\mathcal{B}$ is an algebra, called the ambichiral algebra. The modular invariant is defined by:
\begin{equation}
\mathcal{M}_{\lambda\mu}= \sum_{c \in \mathcal{B}} (E_{\tilde{0}})_{\lambda c} (E_{\tilde{0}})_{\mu c} \;.
\end{equation}
The extended characters $\hat{\chi}_c^k$, labelled by vertices $c \in \mathcal{B}$, are defined by:
\begin{equation}
\hat{\chi}_{\tilde{\lambda}}^k = \chi_{\lambda}^k + \chi_{A(\lambda)}^k + 
\chi_{A^2(\lambda)}^k \;,
\qquad \qquad \hat{\chi}_{\alpha^{(i)}}^k    = \chi_{\alpha}^k \;.
\end{equation}
The modular invariant partition functions associated to the subgroup orbifold graphs $\mathcal{D}_{k\equiv 0}$ are block-diagonal with respect to the extended characters:
\begin{equation}
\mathcal{Z}(\mathcal{D}_{k\equiv 0}) = \sum_{c \in \mathcal{B}} |\hat{\chi}_c^k|^2 =
\frac{1}{3} \sum_{\tiny \begin{array}{c} \lambda \in \mathcal{P}_+^k \\ t(\lambda)=0 
\end{array}} |\chi_{\lambda}^k + \chi_{A(\lambda)}^k+\chi_{A^2(\lambda)}^k|^2 
\qquad \qquad \textrm{for } k = 0 \mod 3 \;.
\end{equation}

\subsection{The conjugated $\mathcal{D}_{k\equiv 0}$ series}
Conjugation is defined on the $\mathcal{A}_k$ graphs: $C(\lambda)=C(\lambda_1,\lambda_2)
=(\lambda_2,\lambda_1)=\lambda^*$. From the induction rules $\mathcal{D}_k \hookleftarrow \mathcal{A}_k$, conjugation is also well defined on the $\mathcal{D}_{k\equiv 0}$ graphs. The vertices $\alpha^{(i)}$ of $\mathcal{D}_{k\equiv 0}$ coming from the induction of the central vertex $\alpha=(\frac{k}{3},\frac{k}{3})$ of $\mathcal{A}_k$ 
are self-conjugated: $C(\alpha^{(i)})=\alpha^{(i)}$. Recall that vertices 
$\tilde{\lambda}$ come from the induction $\tilde{\lambda} \hookleftarrow \lbrace \lambda, A(\lambda),A^2(\lambda) \rbrace$. A straightforward calculation shows that:
\begin{equation}
A(\lambda^*) = A^2(\lambda)^* \;, \qquad \qquad \qquad A^2(\lambda^*) = A(\lambda)^* \;.
\label{conjdk}
\end{equation}
If $\lambda = \lambda^*$, then $A(\lambda)=A^2(\lambda)^*$: the vertex $\tilde{\lambda}$ of $\mathcal{D}_{k\equiv 0}$ is self-conjugate: $\tilde{\lambda}=\tilde{\lambda}^*$.
If $\lambda \not= \lambda^*$, then the conjugation on elements of the orbit of $\lambda$ give the elements of the orbits of $\lambda^*$. Therefore, conjugation is well-defined on all vertices of $\mathcal{D}_{k\equiv 0}$. The modular invariant corresponding to the conjugated 
$\mathcal{D}_{k\equiv 0}$ model is $\mathcal{M}_{\lambda\mu}= \sum_{c \in \mathcal{B}} (E_{\tilde{0}})_{\lambda c} (E_{\tilde{0}})_{\mu c^*}$. To this modular invariant is associated the conjugated orbifold graphs, denoted $\mathcal{D}_{k\equiv 0}^*$, that are module graphs. 
Once again we recall that we do not analyze these module graphs $\mathcal{D}_{k\equiv 0}^*$, the associated partition function is obtained from a conjugation defined directly on the $\mathcal{D}_{k\equiv 0}$ graphs. The corresponding modular invariant partition functions are given by:
\begin{eqnarray}
\nonumber \mathcal{Z}(\mathcal{D}_{k\equiv 0}^*) &=& \sum_{c \in \mathcal{B}} \hat{\chi}_c^k 
\overline{\hat{\chi}_{c^*}^k} \qquad \qquad \qquad \qquad \qquad \textrm{for } k = 0 \mod 3 \\
{ } &=& 
\frac{1}{3} \sum_{\tiny \begin{array}{c} \lambda \in \mathcal{P}_+^k \\ t(\lambda)=0 
\end{array}} (\chi_{\lambda}^k + \chi_{A(\lambda)}^k+\chi_{A^2(\lambda)}^k ) \;
(\overline{\chi_{\lambda^*}^k} + \overline{\chi_{A(\lambda)^*}^k} + 
\overline{\chi_{A^2(\lambda)^*}^k} ) 
 \;.
\end{eqnarray}
Notice that for $k=3$ and 6, the vertices belonging to the ambichiral subset $\mathcal{B}$ are self-conjugate. Therefore, the modular invariant partition function of the conjugated model $\mathcal{D}_k^*$ is equal to the usual one: $\mathcal{Z}(\mathcal{D}_3^*)=\mathcal{Z}(\mathcal{D}_3)$,
$\mathcal{Z}(\mathcal{D}_6^*)=\mathcal{Z}(\mathcal{D}_6)$, even if the associated graphs are different: $\mathcal{D}_3^* \not= \mathcal{D}_3$, $\mathcal{D}_6^* \not= \mathcal{D}_6$.

\subsection{The $\mathcal{D}_3 = \mathcal{A}_3/3$ orbifold graph}
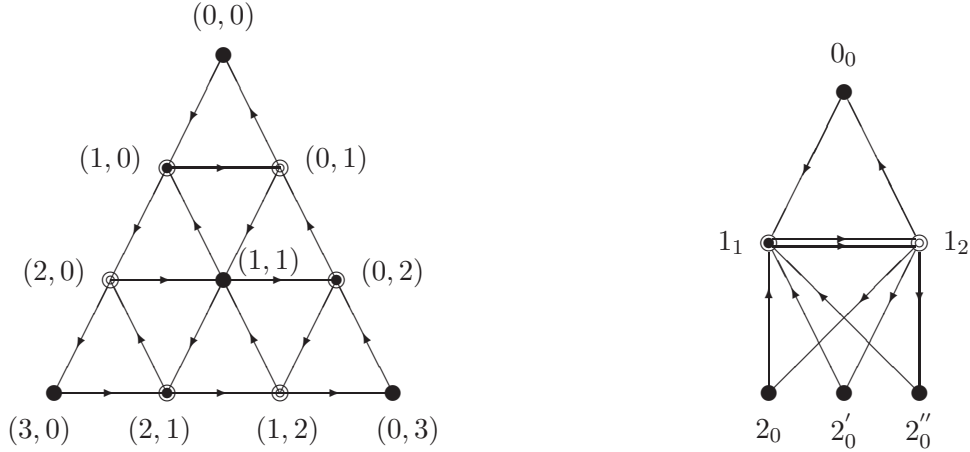
\begin{figure}[h]
\setlength{\unitlength}{1mm}
\begin{center}
\begin{picture}(150,61)

\put(20,10){\begin{picture}(15,15)
\put(0,0){\line(1,0){15}}  
\put(0,0){\line(1,2){7.5}}  
\put(15,0){\line(-1,2){7.5}}
\put(7.5,0){\vector(1,0){0.5}}
\put(11.5,7){\vector(-1,2){0,5}}
\put(3.5,7){\vector(-1,-2){0,5}}
\end{picture}}

\put(35,10){\begin{picture}(15,15)
\put(0,0){\line(1,0){15}}  
\put(0,0){\line(1,2){7.5}}  
\put(15,0){\line(-1,2){7.5}}
\put(7.5,0){\vector(1,0){0.5}}
\put(11.5,7){\vector(-1,2){0,5}}
\put(3.5,7){\vector(-1,-2){0,5}}
\end{picture}}

\put(50,10){\begin{picture}(15,15)
\put(0,0){\line(1,0){15}}  
\put(0,0){\line(1,2){7.5}}  
\put(15,0){\line(-1,2){7.5}}
\put(7.5,0){\vector(1,0){0.5}}
\put(11.5,7){\vector(-1,2){0,5}}
\put(3.5,7){\vector(-1,-2){0,5}}
\end{picture}}

\put(27.5,25){\begin{picture}(15,15)
\put(0,0){\line(1,0){15}}  
\put(0,0){\line(1,2){7.5}}  
\put(15,0){\line(-1,2){7.5}}
\put(7.5,0){\vector(1,0){0.5}}
\put(11.5,7){\vector(-1,2){0,5}}
\put(3.5,7){\vector(-1,-2){0,5}}
\end{picture}}

\put(42.5,25){\begin{picture}(15,15)
\put(0,0){\line(1,0){15}}  
\put(0,0){\line(1,2){7.5}}  
\put(15,0){\line(-1,2){7.5}}
\put(7.5,0){\vector(1,0){0.5}}
\put(11.5,7){\vector(-1,2){0,5}}
\put(3.5,7){\vector(-1,-2){0,5}}
\end{picture}}

\put(35,40){\begin{picture}(15,15)
\put(0,0){\line(1,0){15}}  
\put(0,0){\line(1,2){7.5}}  
\put(15,0){\line(-1,2){7.5}}
\put(7.5,0){\vector(1,0){0.5}}
\put(11.5,7){\vector(-1,2){0,5}}
\put(3.5,7){\vector(-1,-2){0,5}}
\end{picture}}

\put(20,10){\circle*{2}}
\put(65,10){\circle*{2}}
\put(42.5,25){\circle*{2}}
\put(42.5,55){\circle*{2}}
\put(35,10){\circle*{1.5}}
\put(35,10){\circle{2}}
\put(57.5,25){\circle*{1.5}}
\put(57.5,25){\circle{2}}
\put(35,40){\circle*{1.5}}
\put(35,40){\circle{2}}
\put(50,10){\circle{2}}
\put(50,10){\circle{1}}
\put(27.5,25){\circle{2}}
\put(27.5,25){\circle{1}}
\put(50,40){\circle{2}}
\put(50,40){\circle{1}}

\put(18,5){\makebox(0,0){$(3,0)$}}  
\put(34,5){\makebox(0,0){$(2,1)$}}  
\put(51,5){\makebox(0,0){$(1,2)$}}  
\put(67,5){\makebox(0,0){$(0,3)$}}  
\put(20,26){\makebox(0,0){$(2,0)$}}  
\put(65,26){\makebox(0,0){$(0,2)$}}  
\put(27.5,41){\makebox(0,0){$(1,0)$}}  
\put(57.5,41){\makebox(0,0){$(0,1)$}}  
\put(42.5,60){\makebox(0,0){$(0,0)$}}  
\put(48.5,27.5){\makebox(0,0){$(1,1)$}}  
 
\put(105,0){\begin{picture}(40,60)
\put(10.5,29.5){\line(1,0){19}}
\put(10.5,30.5){\line(1,0){19}}
\put(20,29.5){\vector(1,0){0.5}}
\put(20,30.5){\vector(1,0){0.5}}
\put(29.5,31){\line(-1,2){9}}
\put(29.5,31){\vector(-1,2){5}}
\put(19.5,49){\line(-1,-2){9}}
\put(19.5,49){\vector(-1,-2){5}}
\put(10,10){\line(0,1){18.75}}
\put(30,10){\line(0,1){18.75}}
\put(10,10){\line(1,1){19.25}}
\put(30,10){\line(-1,1){19.25}}
\put(20,10){\line(1,2){9.5}}
\put(20,10){\line(-1,2){9.5}}
\put(10,23.5){\vector(0,1){0.5}}
\put(30,22.5){\vector(0,-1){0.5}}
\put(22.5,22.5){\vector(-1,-1){0.5}}
\put(17,23){\vector(-1,1){0.5}}
\put(26.5,23){\vector(-1,-2){0.5}}
\put(13.5,23){\vector(-1,2){0.5}}

\put(20,50){\circle*{2}}
\put(10,10){\circle*{2}}
\put(20,10){\circle*{2}}
\put(30,10){\circle*{2}}
\put(10,30){\circle*{1.25}}
\put(10,30){\circle{2}}
\put(30,30){\circle{1}}
\put(30,30){\circle{2}}

\put(20,55){\makebox(0,0){$0_0$}}  
\put(5,30){\makebox(0,0){$1_1$}}  
\put(35,30){\makebox(0,0){$1_2$}}  
\put(10,5){\makebox(0,0){$2_0$}}  
\put(20,5){\makebox(0,0){$2_0^{'}$}}  
\put(30,5){\makebox(0,0){$2_0^{''}$}}  
\end{picture}}

\end{picture}
\end{center}
\caption{The $\mathcal{A}_{3}$ graph and its orbifold graph $\mathcal{D}_3 = \mathcal{A}_3/3$.}
\label{fig:grapha3d3}
\end{figure}

Let us analyze the $k=3$ example. The $\mathcal{A}_3$ graph has 10 vertices labelled by irreps of $\widehat{su}(3)_3$, that we choose in the following order:
(0,0), (1,0), (0,1), (2,0), (1,1), (0,2), (3,0), (2,1), (1,2), (0,3). 
Its level is $k=3$ and its altitude (generalized Coxeter number) is $\kappa =6$. The $\mathcal{A}_3$ graph is displayed at the left hand side of Figure \ref{fig:grapha3d3}.

\noindent The action of the $\ZZ_{3}-$automorphism $A$ on the vertices of $A_{3}$ is given by:
\begin{equation}\label{vtxA33}
\begin{array}{ccccccc}
A(0,0)=(3,0) &\quad& A(1,0)=(2,1) &\quad& A(0,1)=(2,0) &\quad&   \\
A(3,0)=(0,3) &\quad& A(2,1)=(0,2) &\quad& A(2,0)=(1,2) &\quad& A(1,1)=(1,1)\\  
A(0,3)=(0,0) &\quad& A(0,2)=(1,0) &\quad& A(1,2)=(0,1) &\quad&   
\end{array}
\end{equation}
We have 4 independent orbits $\mathcal{O}_{A}(\lambda)$, but one of them is special since it is made of a unique fixed point: the vertex $(1,1)$ is $A-$ invariant, so this vertex has to be 3-plicated in the orbifold graph. The orbifold graph possesses 6 vertices with well defined triality $\tau$:
\begin{equation}
\begin{array}{ccrcccc}
\mathcal{O}_{A} (0,0) &\rightarrow& 0_0 \in \mathcal{A}_3/3 &\qquad& \textrm{unity} &\quad& \tau=0 \\
\mathcal{O}_{A} (1,0) &\rightarrow& 1_1 \in \mathcal{A}_3/3 &\qquad& \textrm{left generator} &\quad& \tau=1 \\
\mathcal{O}_{A} (0,1) &\rightarrow& 1_2 \in \mathcal{A}_3/3 &\qquad& \textrm{right generator} &\quad& \tau=2 \\
\mathcal{O}_{A} (1,1) &\rightarrow& \quad 2_0, 2_0',2_0'' \in \mathcal{A}_3/3 &\qquad&  &\quad& \tau=0
\end{array}
\end{equation}
The adjacency matrix of the orbifold graph (for the left generator $1_1$) is obtained from
the one of the $A_3$ graph by Equations (\ref{adjmatrorbifold}). We choose the following order for the vertices of $\mathcal{D}_3$: $\lbrace 0_0, 1_1, 1_2, 2_0, 2_0', 2_0'' \rbrace$. The adjacency matrices are given by:
\begin{equation}
G_{(1,0)}^{\mathcal{A}_{3}}= 
\begin{pmatrix}
. & 1 & . & . & . & . & . & . & . & . \\
. & . & 1 & 1 & . & . & . & . & . & . \\
1 & . & . & . & 1 & . & . & . & . & . \\
. & . & . & . & 1 & . & 1 & . & . & . \\
. & 1 & . & . & . & 1 & . & 1 & . & . \\
. & . & 1 & . & . & . & . & . & 1 & . \\
. & . & . & . & . & . & . & 1 & . & . \\
. & . & . & 1 & . & . & . & . & 1 & . \\
. & . & . & . & 1 & . & . & . & . & 1 \\
. & . & . & . & . & 1 & . & . & . & . 
\end{pmatrix}
\qquad \qquad
G_{1_1}^{\mathcal{D}_3} = 
\begin{pmatrix}
  . & 1 & . & . & . & . \\
  . & . & 2 & . & . & . \\
  1 & . & . & 1 & 1 & 1 \\
  . & 1 & . & . & . & . \\
  . & 1 & . & . & . & . \\
  . & 1 & . & . & . & .
\end{pmatrix}
\end{equation}
The corresponding graph $\mathcal{D}_3 = \mathcal{A}_{3}/3$ is displayed at the right hand side
of Figure \ref{fig:grapha3d3}. 
The set of exponents of $\mathcal{D}_3$ is: $Exp (\mathcal{D}_3)
=\left\{ (0,0) , (3,0) , (0,3) , (1,1) , (1,1) , (1,1) \right\}$, which is a
subset of $Exp\left( \mathcal{A}_{3}\right) $. The spectrum of $G^{\mathcal{D}_3}$ is then:
\begin{equation}\label{specA33}
Spec (\mathcal{D}_3) =\left\{ \gamma^{(0,0)}=2 ; \gamma^{(3,0)}
=\left( \gamma^{(0,3)} \right)^{\ast} = 2 \exp \left( 2i\pi /3\right) ;\gamma^{(1,1)}
=0 \textrm{ (triple)} \right\} \;.
\end{equation}
The norm of the graph is given by $\beta =\gamma^{(0,0} = [3]_q = 1+2\cos(2\pi/\kappa)=2$.

\paragraph{Multiplication table of the fusion graph algebra:}
A faithfull representation of the fusion graph algebra of $\mathcal{D}_3$ 
is provided by the fusion matrices $G_a$, $a \in \mathcal{D}_3$. They are $6 \times 6$ square matrices subject to the following polynomials:
\begin{equation}\label{mtxA33}
G_{0_0}=\munite_{6}, \qquad G_{1_{1}} = G^{\mathcal{D}_3}, \qquad G_{1_{2}} = \left[
G_{1_1} \right]^{T}, \qquad G_{2_{0}} + G_{2_0'}+G_{2_0''} = G_{1_{1}} \cdot G_{1_{1}} - G_{0} \;.
\end{equation}
The vertices $2_0,2_0'$ and $2_0''$ of $\mathcal{D}_3$ come from the split of the
vertex $(1,1)$ of $\mathcal{A}_3$, which corresponds to a real (self-conjugated) irrep of $\widehat{su}(3)_3$ . So the vertices $2_0,2_0'$ and $2_0''$ are also self-conjugated,
and the corresponding $G_a$ fusion matrices are symmetric. Imposing the integrality condition 
on their coefficients, we get a unique solution: 
\begin{equation*}
G_{2_0} = 
\begin{pmatrix}
. & . & . & 1 & . & . \\
. & 1 & . & . & . & . \\
. & . & 1 & . & . & . \\
1 & . & . & . & . & . \\
. & . & . & . & . & 1 \\
. & . & . & . & 1 & . 
\end{pmatrix}
\qquad
G_{2_0'} = 
\begin{pmatrix}
. & . & . & . & 1 & . \\
. & 1 & . & . & . & . \\
. & . & 1 & . & . & . \\
. & . & . & . & . & 1 \\
1 & . & . & . & . & . \\
. & . & . & 1 & . & . 
\end{pmatrix}
\qquad
G_{2_0''} = 
\begin{pmatrix}
. & . & . & . & . & 1 \\
. & 1 & . & . & . & . \\
. & . & 1 & . & . & . \\
. & . & . & . & 1 & . \\
. & . & . & 1 & . & . \\
1 & . & . & . & . & . 
\end{pmatrix}
\end{equation*}
We can now complete the multiplication table of the graph algebra of $\mathcal{D}_3$.
\begin{table}[h]\label{tabA33}
\begin{center}
\begin{tabular}{|c||c|c|c|c|c|c|}
\hline 
$\cdot $ & $0_0$ & $2_{0}$ & $2_{0}'$ & $2_{0}''$ & $1_{1}$ & $1_{2}$ \\ 
\hline 
\hline
$0_0$ & $0_0$ & $2_{0}$ & $2_{0}'$ & $2_{0}''$ & $1_{1}$ & $1_{2}$ \\ 
\hline 
$2_{0}$ & $2_0$ & $0_0$ & $2_{0}''$ & $2_{0}'$ & $1_{1}$ & $1_{2}$ \\ 
\hline 
$2_{0}'$ & $2_{0}'$ & $2_{0}''$ & $0_{0}$ & $2_0$ & $1_{1}$ & $1_{2}$ \\ 
\hline 
$2_{0}''$ & $2_{0}''$ &$2_{0}$ & $0_0$ & $2_{0}$ & $1_{1}$ & $1_{2}$ \\ 
\hline 
$1_{1}$ & $1_{1}$ & $1_{1}$ & $1_{1}$ & $1_{1}$ & $1_2 + 1_2$ & $0_0 +2_{0}+2_{0}'+2_{0}''$ \\ 
\hline 
$1_{2}$ & $1_{2}$ & $1_{2}$ & $1_{2}$ & $1_{2}$ & $0_0 + 2_{0}+2_{0}'+2_{0}''$ & $ 1_1 +1_1$ \\
\hline
\end{tabular}
\end{center}
\caption{Multiplication table for the graph algebra of $\mathcal{D}_3$.}
\end{table}

\paragraph{Branching rules and the modular invariant partition function}
The action of vertices $\lambda$ of $\mathcal{A}_{3}$ over
vertices $a$ of $\mathcal{D}_{3}$ is encoded by the set of 
$10$ fused matrices $F_{\lambda}$, obtained by the $SU(3)$ truncated recursion formula
(\ref{su3recform}), with initial conditions $F_{(1,0)}=G_{1_1}$ and $F_{(0,1)}=G_{1_2}$. 
They are explicitly given by: 
\begin{equation}\label{FusA33}
\begin{array}{lcl}
F_{(0,0)}=F_{(3,0)}=F_{(0,3)}=\munite_{6}  &\qquad& 
F_{(1,0)}=F_{(0,2)}=F_{(2,1)}= G_{1_1} \\
F_{(0,1)}=F_{(2,0)}=F_{(1,2)}= G_{1_2}     &\qquad&
F_{(1,1)}=G_{2_{0}}+G_{2_{0}'}+G_{2_{0}''} 
\end{array}
\end{equation}
The essential matrix $E_{0_0}$ is defined by $(E_{0_0})_{\lambda a} = (F_{\lambda})_{0_0 a}$. 
It is a rectangular matrix with 10 lines labelled by vertices $\lambda \in \mathcal{A}_3$ and 6 
columns labelled by vertices $a \in \mathcal{D}_3$. This matrix encodes the 
branching rules $ \mathcal{A}_3 \hookrightarrow \mathcal{D}_3$ (lines of $E_{0_0}$):
\begin{equation}\label{EssA33}
E_{0_0}(\mathcal{D}_{3}) =
\begin{pmatrix}
1 & 0 & 0 & 0 & 0 & 0 \\
0 & 1 & 0 & 0 & 0 & 0 \\
0 & 0 & 1 & 0 & 0 & 0 \\
0 & 0 & 1 & 0 & 0 & 0 \\
0 & 0 & 0 & 1 & 1 & 1 \\
0 & 1 & 0 & 0 & 0 & 0 \\
1 & 0 & 0 & 0 & 0 & 0 \\
0 & 1 & 0 & 0 & 0 & 0 \\
0 & 0 & 1 & 0 & 0 & 0 \\
1 & 0 & 0 & 0 & 0 & 0 
\end{pmatrix}
\qquad \qquad  \qquad \qquad 
\begin{array}{rcl}
(0,0) &\hookrightarrow& 0_0 \\
(1,0) &\hookrightarrow& 1_1 \\
(0,1) &\hookrightarrow& 1_2 \\
(2,0) &\hookrightarrow& 1_2 \\
(1,1) &\hookrightarrow& 2_0 + 2_0' + 2_0'' \\
(0,2) &\hookrightarrow& 1_1 \\
(3,0) &\hookrightarrow& 0_0 \\
(2,1) &\hookrightarrow& 1_1 \\
(1,2) &\hookrightarrow& 1_2 \\
(0,3) &\hookrightarrow& 0_0 
\end{array}
\end{equation}
as well as the induction rules $ \mathcal{D}_3 \hookleftarrow \mathcal{A}_3$ (columns of $E_{0_0}$):
\begin{equation}
\begin{array}{rclcrcl}
0_0 &\hookleftarrow& (0,0) \,,\, (3,0) \,,\, (0,3)  &\qquad& 2_0 &\hookleftarrow& (1,1) \\
1_1 &\hookleftarrow& (1,0) \,,\, (0,2) \,,\, (2,1)  &\qquad& 2_0' &\hookleftarrow& (1,1)  \\
1_2 &\hookleftarrow& (0,1) \,,\, (2,0) \,,\, (1,2)  &\qquad& 2_0'' &\hookleftarrow& (1,1)
\end{array}
\end{equation}
The values of the modular operator $\hat{T}= (\lambda_1+1)^2 + (\lambda_1+1)(\lambda_2+1)
+(\lambda_2+1)^2  \mod 18$ on the vertices $\lambda=(\lambda_1,\lambda_2)$ of $\mathcal{A}_3$ is given in the next table.
\begin{table}[hhh] 
$$ 
\begin{array}{|c||c|c|c|c|c|c|c|} 
\hline 
(\lambda_1,\lambda_2) & (0,0) & (1,0) &(2,0) &(3,0) &(2,1)  &(1,1)  \\ 
                      &  {}   & (0,1) &(0,2) &(0,3) &(1,2)  &       \\ 
\hline 
\hline 
\hat{T}  & 15 & 1 & 7 & 15 & 13 & 6  \\ 
\hline 
\end{array} 
$$ 
\caption{Modular operator values $\hat{T}$ on vertices of the $\mathcal{A}_3$ graph.}
\end{table}

\noindent We see that the only vertices $a \in \mathcal{D}_3$ for which T is well defined is  the set $\mathcal{B} = \{0_0, 2_0, 2_0',2_0''\}$. We can check that $\mathcal{B}$ is a subalgebra of the graph algebra of $\mathcal{D}_3$, and its
complement spanned by $\left\{ 1_{1},\;1_{2}\right\} $ is
$\mathcal{B}-$invariant.  We therefore have a characterization of the ambichiral algebra $\mathcal{B}$ by modular properties. The modular invariant $\mathcal{M}$ is obtained by:
\begin{equation}
\mathcal{M}_{\lambda\mu} = \sum_{c \in \mathcal{B}} (F_{\lambda})_{0_0\,c} (F_{\mu})_{0_0\,c} = (E_{0_0})\,(E_{0_0}^{red})^{tr} \;,
\end{equation}
where $E_{0_0}^{red}$ is the reduced essential matrix obtained from $E_{0_0}$ by setting to zero the entries of columns labelled by $1_{1}$ and $1_{2}$. We get:
\begin{equation}\label{torA33}
\mathcal{M}(\mathcal{D}_{3}) 
=
\begin{pmatrix}
1 & \cdot & \cdot & \cdot & \cdot & \cdot & 1 & \cdot & \cdot & 1 \\
\cdot & \cdot & \cdot & \cdot & \cdot & \cdot & \cdot & \cdot & \cdot & \cdot \\
\cdot & \cdot & \cdot & \cdot & \cdot & \cdot & \cdot & \cdot & \cdot & \cdot \\
\cdot & \cdot & \cdot & \cdot & \cdot & \cdot & \cdot & \cdot & \cdot & \cdot \\
\cdot & \cdot & \cdot & \cdot & 3 & \cdot & \cdot & \cdot & \cdot & \cdot \\
\cdot & \cdot & \cdot & \cdot & \cdot & \cdot & \cdot & \cdot & \cdot & \cdot \\
1 & \cdot & \cdot & \cdot & \cdot & \cdot & 1 & \cdot & \cdot & 1 \\
\cdot & \cdot & \cdot & \cdot & \cdot & \cdot & \cdot & \cdot & \cdot & \cdot \\
\cdot & \cdot & \cdot & \cdot & \cdot & \cdot & \cdot & \cdot & \cdot & \cdot \\
1 & \cdot & \cdot & \cdot & \cdot & \cdot & 1 & \cdot & \cdot & 1
\end{pmatrix}
\end{equation}
The corresponding modular invariant partition function $\mathcal{Z}(\mathcal{D}_3)$ reads:
\begin{equation}\label{mpfA33}
Z(\mathcal{D}_{3}) = \mid \chi _{\left( 0,0\right) }^3+\chi
_{\left( 3,0\right) }^3+\chi _{\left( 0,3\right) }^3\mid ^{2}+3\mid
\chi _{\left( 1,1\right) }^3\mid ^{2} \;.
\end{equation}


\subsection{The orbifold graph $\mathcal{D}_9$ and the exceptional twist $\xi$}

\begin{figure}[h]
\setlength{\unitlength}{1.0mm}
\begin{center}
\begin{picture}(140,115)

\put(5,5){\begin{picture}(60,105)
\put(0,45){\begin{picture}(60,60)

\put(0,0){\begin{picture}(15,15)
\put(0,0){\line(1,0){15}}  
\put(0,0){\line(1,2){7.5}}  
\put(15,0){\line(-1,2){7.5}}
\put(7.5,0){\vector(1,0){0.5}}
\put(11.5,7){\vector(-1,2){0,5}}
\put(3.5,7){\vector(-1,-2){0,5}}
\end{picture}}

\put(15,0){\begin{picture}(15,15)
\put(0,0){\line(1,0){15}}  
\put(0,0){\line(1,2){7.5}}  
\put(15,0){\line(-1,2){7.5}}
\put(7.5,0){\vector(1,0){0.5}}
\put(11.5,7){\vector(-1,2){0,5}}
\put(3.5,7){\vector(-1,-2){0,5}}
\end{picture}}

\put(30,0){\begin{picture}(15,15)
\put(0,0){\line(1,0){15}}  
\put(0,0){\line(1,2){7.5}}  
\put(15,0){\line(-1,2){7.5}}
\put(7.5,0){\vector(1,0){0.5}}
\put(11.5,7){\vector(-1,2){0,5}}
\put(3.5,7){\vector(-1,-2){0,5}}
\end{picture}}

\put(45,0){\begin{picture}(15,15)
\put(0,0){\line(1,0){15}}  
\put(0,0){\line(1,2){7.5}}  
\put(15,0){\line(-1,2){7.5}}
\put(7.5,0){\vector(1,0){0.5}}
\put(11.5,7){\vector(-1,2){0,5}}
\put(3.5,7){\vector(-1,-2){0,5}}
\end{picture}}

\put(7.5,15){\begin{picture}(15,15)
\put(0,0){\line(1,0){15}}  
\put(0,0){\line(1,2){7.5}}  
\put(15,0){\line(-1,2){7.5}}
\put(7.5,0){\vector(1,0){0.5}}
\put(11.5,7){\vector(-1,2){0,5}}
\put(3.5,7){\vector(-1,-2){0,5}}
\end{picture}}

\put(22.5,15){\begin{picture}(15,15)
\put(0,0){\line(1,0){15}}  
\put(0,0){\line(1,2){7.5}}  
\put(15,0){\line(-1,2){7.5}}
\put(7.5,0){\vector(1,0){0.5}}
\put(11.5,7){\vector(-1,2){0,5}}
\put(3.5,7){\vector(-1,-2){0,5}}
\end{picture}}

\put(37.5,15){\begin{picture}(15,15)
\put(0,0){\line(1,0){15}}  
\put(0,0){\line(1,2){7.5}}  
\put(15,0){\line(-1,2){7.5}}
\put(7.5,0){\vector(1,0){0.5}}
\put(11.5,7){\vector(-1,2){0,5}}
\put(3.5,7){\vector(-1,-2){0,5}}
\end{picture}}

\put(15,30){\begin{picture}(15,15)
\put(0,0){\line(1,0){15}}  
\put(0,0){\line(1,2){7.5}}  
\put(15,0){\line(-1,2){7.5}}
\put(7.5,0){\vector(1,0){0.5}}
\put(11.5,7){\vector(-1,2){0,5}}
\put(3.5,7){\vector(-1,-2){0,5}}
\end{picture}}

\put(30,30){\begin{picture}(15,15)
\put(0,0){\line(1,0){15}}  
\put(0,0){\line(1,2){7.5}}  
\put(15,0){\line(-1,2){7.5}}
\put(7.5,0){\vector(1,0){0.5}}
\put(11.5,7){\vector(-1,2){0,5}}
\put(3.5,7){\vector(-1,-2){0,5}}
\end{picture}}

\put(22.5,45){\begin{picture}(15,15)
\put(0,0){\line(1,0){15}}  
\put(0,0){\line(1,2){7.5}}  
\put(15,0){\line(-1,2){7.5}}
\put(7.5,0){\vector(1,0){0.5}}
\put(11.5,7){\vector(-1,2){0,5}}
\put(3.5,7){\vector(-1,-2){0,5}}
\end{picture}}

\put(30,0){\circle*{2}}
\put(30,30){\circle*{2}}
\put(30,60){\circle*{2}}
\put(7.5,15){\circle*{2}}
\put(52.5,15){\circle*{2}}

\put(0,0){\circle*{1.5}}
\put(0,0){\circle{2}}
\put(45,0){\circle*{1.5}}
\put(45,0){\circle{2}}
\put(22.5,15){\circle*{1.5}}
\put(22.5,15){\circle{2}}
\put(45,30){\circle*{1.5}}
\put(45,30){\circle{2}}
\put(22.5,45){\circle*{1.5}}
\put(22.5,45){\circle{2}}

\put(15,0){\circle{2}}
\put(15,0){\circle{1}}
\put(60,0){\circle{2}}
\put(60,0){\circle{1}}
\put(15,30){\circle{2}}
\put(15,30){\circle{1}}
\put(37.5,15){\circle{2}}
\put(37.5,15){\circle{1}}
\put(37.5,45){\circle{2}}
\put(37.5,45){\circle{1}}
\end{picture}}

\put(15,0){\circle*{2}}
\put(30,0){\circle*{2}}
\put(45,0){\circle*{2}}
\put(30,30){\circle*{2}}
\put(15,15){\circle*{1.5}}
\put(15,15){\circle{2}}
\put(45,15){\circle{2}}
\put(45,15){\circle{1}}

\put(15.5,14.5){\line(1,0){29}}
\put(15.5,15.5){\line(1,0){29}}
\put(15,15){\line(1,2){15}}
\put(45,15){\line(-1,2){15}}

\put(30,14.5){\vector(1,0){0.5}}
\put(30,15.5){\vector(1,0){0.5}}
\put(22,29){\vector(-1,-2){0}}
\put(37.5,30){\vector(-1,2){0.5}}

\put(15,0.5){\line(0,1){14}}
\put(45,0.5){\line(0,1){14}}
\put(15,7.5){\vector(0,1){1}}
\put(45,7.5){\vector(0,-1){1}}

\put(30,0){\line(1,1){14.2}}
\put(30,0){\line(-1,1){14.2}}
\put(30,0){\vector(-1,1){8}}
\put(45,15){\vector(-1,-1){8}}

\put(45,0){\line(-2,1){29}}
\put(15,0){\line(2,1){29}}
\put(45,0){\vector(-2,1){20}}
\put(44,14.5){\vector(-2,-1){10.5}}

\put(15,16){\line(0,1){28}}
\put(45,16){\line(0,1){28}}
\put(15.7,15.7){\line(1,1){29}}
\put(44.3,15.7){\line(-1,1){29}}
\put(15,30){\vector(0,1){0.5}}
\put(45,30){\vector(0,-1){0.5}}

\put(30,30){\vector(1,1){8}}
\put(30,30){\vector(-1,-1){8}}
\put(44.4,15.6){\vector(-1,1){7.5}}
\put(15.6,44.4){\vector(1,-1){7.5}}

\put(30,30){\line(-2,1){29}}
\put(30,30){\line(2,1){29}}
\put(30,30){\vector(-2,1){23}}
\put(59.4,44.7){\vector(-2,-1){8}}

\qbezier(0,45)(30,35)(60,45)
\put(30,40){\vector(1,0){0.5}}

\put(30,108.5){\makebox(0,0){$0_0$}}
\put(18,91){\makebox(0,0){$1_1$}}
\put(42,91){\makebox(0,0){$1_2$}}
\put(10.5,76){\makebox(0,0){$2_1$}}
\put(49.5,76){\makebox(0,0){$2_2$}}
\put(3,61){\makebox(0,0){$3_0$}}
\put(57,61){\makebox(0,0){$3_0^{\prime}$}}
\put(-4.5,45){\makebox(0,0){$4_1$}}
\put(64.5,45){\makebox(0,0){$4_2$}}
\put(30,80){\makebox(0,0){$2_0$}}
\put(22.5,65){\makebox(0,0){$3_1$}}
\put(37.5,65){\makebox(0,0){$3_2$}}
\put(30,50){\makebox(0,0){$4_0$}}
\put(15,50){\makebox(0,0){$4_2^{\prime}$}}
\put(45,50){\makebox(0,0){$4_1^{\prime}$}}
\put(30,33.5){\makebox(0,0){$5_0$}}
\put(10.5,15){\makebox(0,0){$5_1$}}
\put(49.5,15){\makebox(0,0){$5_2$}}
\put(15,-4.5){\makebox(0,0){$\alpha^{(1)}_0$}}
\put(30,-4.5){\makebox(0,0){$\alpha^{(2)}_0$}}
\put(45,-4.5){\makebox(0,0){$\alpha^{(3)}_0$}}

\end{picture}}

\put(80,55){$
\begin{array}{rclcc}
\mathcal{B} & & & & \hat{T} \\
 \\
0_0  &\hookleftarrow& (0,0),(9,0),(0,9) &\qquad& 3 \\ 
2_0  &\hookleftarrow& (1,1),(7,1),(1,7) &\qquad& 12 \\ 
3_0  &\hookleftarrow& (3,0),(6,3),(0,6) &\qquad& 21 \\ 
3_0^{\prime}  &\hookleftarrow& (0,3),(6,0),(3,6) &\qquad& 21 \\ 
4_0  &\hookleftarrow& (2,2),(5,2),(2,5) &\qquad& 27 \\ 
5_0  &\hookleftarrow& (4,4),(4,1),(1,4) &\qquad& 3 \\ 
\alpha_0^{(1)}  &\hookleftarrow& (3,3) &\qquad& 12 \\ 
\alpha_0^{(2)}  &\hookleftarrow& (3,3) &\qquad& 12 \\ 
\alpha_0^{(3)}  &\hookleftarrow& (3,3) &\qquad& 12 
\end{array}
$}

\end{picture}

\end{center}
\caption{The orbifold graph $\mathcal{D}_{9}$ and the subset $\mathcal{B}$.}
\label{fig:d9}
\end{figure}
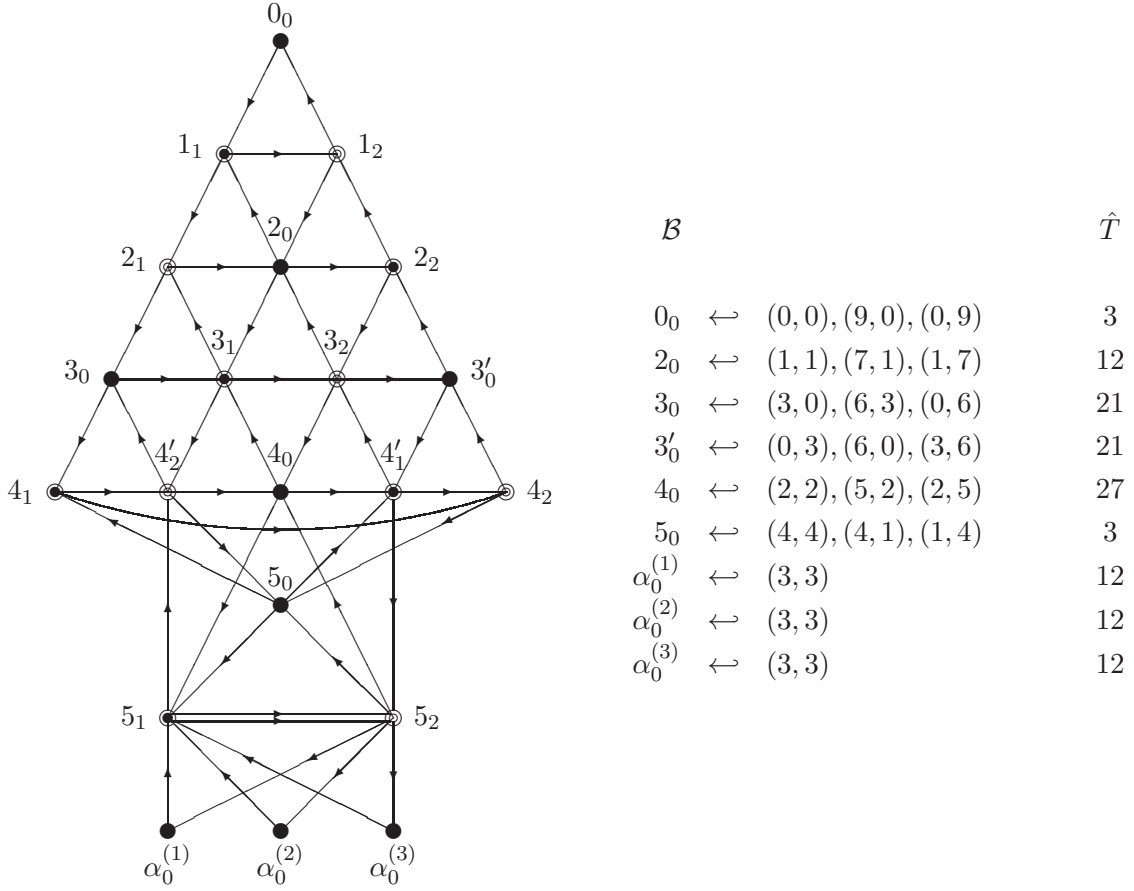

The $\mathcal{D}_9$ graph is displayed in Figure \ref{fig:d9}, it is obtained as the $\mathbb{Z}_3$-orbifold of the  $\mathcal{A}_9$ graph. Triality and conjugation are well-defined. The conjugation corresponds to the symmetry with respect to the vertical axis going through the unit vertex $0_0$, except for the triplicated vertices $\alpha^{(i)}$, wich are self-conjugated (they should better be drawn one upon the other). 

The subset $\mathcal{B}$ of vertices of $\mathcal{D}_9$ with well-defined modular operator value 
$\hat{T}$, determined from the induction mechanism  
$\mathcal{D}_{9} \hookleftarrow \mathcal{A}_9$ and modular properties, is 
$\mathcal{B}=\{0_0,2_0,3_0,3_0^{\prime},4_0,5_0,\alpha^{(1)}_0,\alpha^{(2)}_0,\alpha^{(3)}_0\}$. The extended characters are defined through the induction mechanism (see Figure \ref{fig:d9}). We have for example $\hat{\chi}_{0_0}^9 = \chi_{(0,0)}^9 + \chi_{(9,0)}^9 + \chi_{(0,9)}^9$, 
$\hat{\chi}_{2_0}^9 = \chi_{(1,1)}^9 + \chi_{(7,1)}^9 + \chi_{(1,7)}^9$. The modular invariant partition function associated to the $\mathcal{D}_9$ graph is given by:
\begin{equation}
\mathcal{Z}(\mathcal{D}_9) = \sum_{c \in \mathcal{B}} |\hat{\chi}^9_c|^2 \;.
\end{equation}
Conjugated vertices of $\mathcal{B}$ have the same modular operator value. Elements of $\mathcal{B}$ are self-conjugate except $3_0^*=3_0^{\prime}$. The modular invariant partition function of 
the conjugated $\mathcal{D}_9$ model is given by:
\begin{equation}
\mathcal{Z}(\mathcal{D}_9^*) = \sum_{c \in \mathcal{B}} \hat{\chi}^9_c \, \overline{\hat{\chi}^9_{c^*}} \;.
\end{equation} 
It is associated with the conjugated graph $\mathcal{D}_9^*$, which is a module graph and is not studied in this paper.
At this level $k=9$, we see that the modular operator value of the triplicated vertices is equal to the one 
of another vertex of $\mathcal{B}$, namely $2_0$:  $\hat{T}(\alpha_0^{(i)}) = \hat{T}(2_0)=12$ (see Figure \ref{fig:d9}). This equality leads to the definition of the exceptional twist $\xi$ of $\mathcal{D}_9$:
\begin{equation}
\xi(2_0) = \alpha_0^{(1)}\,, \; \xi(\alpha_0^{(1)}) = 2_0 \,, \qquad \qquad \xi(c)=c \; \textrm{ for } c \in \mathcal{B}, c \not= \{2_0,\alpha_0^{(i)}\}\;.
\end{equation}
This exceptional twist is a $\hat{T}$-invariant involution. The modular invariant partition function of the exceptional twist of $\mathcal{D}_9$ is given by:
\begin{equation}
\begin{array}{rcl}
\mathcal{Z}(\mathcal{D}_9^{\xi}) = \dps \sum_{c \in \mathcal{B}} \hat{\chi}^9_c \, 
\overline{\hat{\chi}^9_{\xi(c)}} &=& |\hat{\chi}_{0_0}^9|^2 
+ |\hat{\chi}_{3_0}^9|^2
+ |\hat{\chi}_{3_0^{\prime}}^9|^2
+ |\hat{\chi}_{4_0}^9|^2
+ |\hat{\chi}_{5_0}^9|^2 \\
&+& |\hat{\chi}_{\alpha_0^{(2)}}^9|^2
+ |\hat{\chi}_{\alpha_0^{(3)}}^9|^2
+ ( \hat{\chi}_{2_0}^9 \, \overline{\hat{\chi}_{\alpha_0^{(1)}}^9} + \textrm{h.c.}) \;.
\end{array}
\end{equation}
It is associated to the exceptional module graph $\mathcal{D}_9^{\xi}$. It is the $SU(3)$
analogue of the $E_7$ graph of the $SU(2)$ series, which is an exceptional twist of the 
$D_{10}$ graph (an orbifold of the $A_{17}$ Dynkin diagram). Combining the conjugation with the exceptional twist $\xi$, we obtain the following modular invariant partition function:
\begin{equation}
\mathcal{Z}(\mathcal{D}_9^{\xi*}) = \sum_{c \in \mathcal{B}} \hat{\chi}^9_c 
\overline{\hat{\chi}_{\xi(c)^*}^9} \;,
\end{equation}
which is associated to the module graph $\mathcal{D}_9^{\xi*}$. Let us emphasize that from the subgroup graph $\mathcal{D}_9$, we obtained, combining the $\hat{T}$-invariant involutions conjugation and the exceptional twist $\xi$, 4 modular invariant partition functions. They are asociated to 4 different Di Francesco-Zuber graphs, 
the subgroup $\mathcal{D}_9$ graph and the 3 modules graphs $\mathcal{D}_9^*$, $\mathcal{D}_9^{\xi}$ and $\mathcal{D}_9^{\xi*}$. We do not need to analyze these module graphs if we are only interessed in obtaining the corresponding modular invariant $\mathcal{Z}$.

\paragraph{Other ``non-physical'' twists}
The values of the modular operator are also equal for the vertices $0_0$ and $5_0$ of $\mathcal{D}_9$. We can therefore define a $\hat{T}$-invariant twist $\psi$ such that $\psi(0_0)=5_0,\psi(5_0)=0_0$ and $\psi(c)=c$. The partition function defined by:
\begin{equation}
\mathcal{Z}(\mathcal{D}_9^{\psi}) = \sum_{c \in \mathcal{B}} \hat{\chi}^9_c \, \overline{\hat{\chi}^9_{\psi(c)}} \;,
\end{equation} 
indeed commutes with the modular matrix $\mathcal{T}$. But it does not commute with the modular 
matrix $\mathcal{S}$, and the corresponding $\mathcal{Z}$ is not modular invariant. As exemplified here, the condition that the twist must be $\hat{T}$-invariant is a necessary but not sufficient condition. Furthermore, in this particular example, we have twisted the unit vertex $0_0$, so the corresponding $\mathcal{M}$ does not satisfy the axiom (P1): 
$\mathcal{M}_{0_0,0_0}\not=1$.

\section{The exceptional cases}
The subgroup graphs $\mathcal{A}_k$ and their orbifold $\mathcal{D}_k=\mathcal{A}_k/3$ have been constructed from primary data coming from CFT fusion rules and orbifold methods. This is not the case for the three exceptional subgroup graphs, namely $\mathcal{E}_5,\mathcal{E}_9$ and $\mathcal{E}_{21}$. They are taken as initial data from \cite{DiF_Zub-graphs} or \cite{Oc-Bariloche}. Most of this section is taken from \cite{Coq_Gil-Tmod} or from the thesis 
\cite{Gil-tesis}. What is new here is the discussion relative to conjugation and/or twist.

\subsection{The $\mathcal{E}_5$ graph} 
\begin{figure}[hhh] 
\begin{center} 
\unitlength 0.30mm 
 
\begin{picture}(180,200)(0,-10) 
 
\put(0,45){\begin{picture}(60,45) 
\put(0,0){\circle{8}} 
\put(0,0){\circle*{6}} 
\put(60,0){\circle{5}} 
\put(60,0){\circle{8}} 
\put(30,45){\circle*{8}} 
\put(0,0){\vector(1,0){32.5}} 
\put(30,0){\line(1,0){30}} 
\put(60,0){\vector(-2,3){16.5}} 
\put(45,22.5){\line(-2,3){15}} 
\put(30,45){\vector(-2,-3){16.5}} 
\put(15,22.5){\line(-2,-3){15}} 
\end{picture}} 
 
\put(120,45){\begin{picture}(60,45) 
\put(0,0){\circle{8}} 
\put(0,0){\circle*{6}} 
\put(60,0){\circle{5}} 
\put(60,0){\circle{8}} 
\put(30,45){\circle*{8}}
\put(0,0){\vector(1,0){32.5}} 
\put(30,0){\line(1,0){30}} 
\put(60,0){\vector(-2,3){16.5}} 
\put(45,22.5){\line(-2,3){15}} 
\put(30,45){\vector(-2,-3){16.5}} 
\put(15,22.5){\line(-2,-3){15}} 
\end{picture}} 
 
\put(60,135){\begin{picture}(60,45) 
\put(0,0){\circle{8}} 
\put(0,0){\circle*{6}} 
\put(60,0){\circle{5}} 
\put(60,0){\circle{8}} 
\put(30,45){\circle*{8}}
\put(0,0){\vector(1,0){32.5}} 
\put(30,0){\line(1,0){30}} 
\put(60,0){\vector(-2,3){16.5}} 
\put(45,22.5){\line(-2,3){15}} 
\put(30,45){\vector(-2,-3){16.5}} 
\put(15,22.5){\line(-2,-3){15}} 
\end{picture}}

\put(0,90){\begin{picture}(60,45) 
\put(0,45){\circle{5}} 
\put(0,45){\circle{8}} 
\put(60,45){\vector(-1,0){32.5}} 
\put(30,45){\line(-1,0){30}} 
\put(30,0){\vector(2,3){16.5}} 
\put(45,22.5){\line(2,3){15}} 
\put(0,45){\vector(2,-3){16.5}} 
\put(15,22.5){\line(2,-3){15}} 
\end{picture}} 
 
\put(120,90){\begin{picture}(60,45) 
\put(60,45){\circle*{6}} 
\put(60,45){\circle{8}} 
\put(60,45){\vector(-1,0){32.5}} 
\put(30,45){\line(-1,0){30}} 
\put(30,0){\vector(2,3){16.5}} 
\put(45,22.5){\line(2,3){15}} 
\put(0,45){\vector(2,-3){16.5}} 
\put(15,22.5){\line(2,-3){15}} 
\end{picture}} 
 
\put(60,0){\begin{picture}(60,45) 
\put(30,0){\circle*{8}} 
\put(60,45){\vector(-1,0){32.5}} 
\put(30,45){\line(-1,0){30}} 
\put(30,0){\vector(2,3){16.5}} 
\put(45,22.5){\line(2,3){15}} 
\put(0,45){\vector(2,-3){16.5}} 
\put(15,22.5){\line(2,-3){15}} 
\end{picture}} 
 
\put(60,135){\vector(0,-1){47.5}} 
\put(60,90){\line(0,-1){45}} 
\put(60,45){\vector(2,1){47.2}} 
\put(105,67.5){\line(2,1){45}} 
\put(150,90){\vector(-2,1){47.2}} 
\put(105,112.5){\line(-2,1){45}} 
 
\put(120,45){\vector(0,1){47.5}} 
\put(120,90){\line(0,1){45}} 
\put(120,135){\vector(-2,-1){47.2}} 
\put(75,112.5){\line(-2,-1){45}} 
\put(30,90){\vector(2,-1){47.2}} 
\put(75,67.5){\line(2,-1){45}}

\put(-12,45){\makebox(0,0){$1_4$}} 
\put(192,45){\makebox(0,0){$1_2$}} 
\put(-12,135){\makebox(0,0){$1_5$}} 
\put(192,135){\makebox(0,0){$1_1$}} 
\put(90,-12){\makebox(0,0){$1_3$}} 
\put(90,192){\makebox(0,0){$1_0$}} 
\put(55,35){\makebox(0,0){$2_5$}} 
\put(125,35){\makebox(0,0){$2_4$}} 
\put(55,145){\makebox(0,0){$2_1$}} 
\put(125,145){\makebox(0,0){$2_2$}} 
\put(20,90){\makebox(0,0){$2_0$}} 
\put(160,90){\makebox(0,0){$2_3$}} 

\end{picture} 
\end{center} 
\caption{The $\mathcal{E}_5$ graph.} 
\label{fig:graphecinq} 
\end{figure}
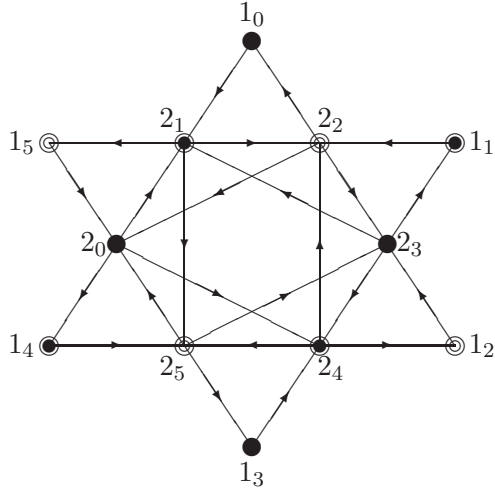

The ${\mathcal E}_5$ graph has 12 vertices (irreps) denoted by 
$1_i, 2_i$, $i =0,1,\ldots,5. $ Its level $k = 5$, its altitude $\kappa = k + 3 = 8$ and its norm $\beta = 1+2\cos(2  \pi /8) = 1 + \sqrt2$. The unity is $1_0$ and the left and right chiral generators (fundamental irreps) are $2_1$ and $2_2$. The graph is 3--colorable and is displayed in Figure \ref{fig:graphecinq}, its orientation corresponding to $2_1$. The graph of the 
${\cal A}$ series with same norm is ${\cal A}_5$, which has $(k+1)(k+2)/2 = 21$ vertices. The eigenvalues of the adjacency matrix of the ${\cal E}_5$ graph are a subset of those of the ${\cal A}_5$ graph. The exponents of ${\cal E}_5$ are:
\begin{equation}
\label{CoexpE5}
\begin{array}{rcl}
(r_1,r_2) &=& (0,0) \;,\;( 2,2) \;,\;( 3,0) \;,\;( 0,3) \;,\;( 5,0) \;,\;(0,5) \\
{ } &{ }& ( 2,1) \;,\; ( 1,2) \;,\;( 2,3) \;,\;( 0,2)  \;,\;( 3,2) \;,\;( 2,0).
\end{array}
\end{equation}

The $\mathcal{E}_5$ graph is a subgroup graph. It determines in a unique way its graph algebra. The commutative multiplication table is given by:
\begin{equation} 
\begin{array}{rcl} 
1_i . 1_j & = & 1_{i+j}    \\ 
1_i . 2_j = 2_i . 1_j & = & 2_{i+j}  \qquad \qquad \qquad \qquad \qquad \qquad \qquad i,j =0,1,\ldots,5 \; \mod 6 \\ 
2_i . 2_j & = & 2_{i+j} + 2_{i+j-3} + 1_{i+j-3} 
\end{array} 
\label{mult_E5} 
\end{equation} 
From this multiplication table we get the fusion matrices $G_a$ associated to the vertices 
$a \in \mathcal{E}_5$. The one corresponding to the generator $2_1$ is the adjacency matrix of the graph.

\paragraph{Induction-restriction} 
The module property of the vector space spanned by the vertices $a$ of the 
$\mathcal{E}_5$ graph under the action of the graph algebra $\mathcal{A}_5$ is coded by the set of 21 fused matrices $F_{\lambda}$ for $\lambda = (\lambda_1,\lambda_2) \in \mathcal{A}_5$. They are obtained by the truncated SU(3) recursion formulae, with starting point $F_{(1,0)}=G_{2_1}$. 
The essential matrix $E_{1_0}$ is obtained from $(E_{1_0})_{\lambda a} = (F_{\lambda})_{1_0 a}$. 
It contains 21 lines labelled by vertices $\lambda \in \mathcal{A}_5$ and 12 columns labelled by vertices $a \in \mathcal{E}_5$. We read from this matrix the branching rules  $\mathcal{A}_5 \hookrightarrow \mathcal{E}_5$ and the induction rules given by: 
$$ 
\begin{array}{ccc} 
1_0 \hookleftarrow (0,0) , (2,2) &\qquad \qquad \qquad&  
2_0 \hookleftarrow (1,1) , (3,0) , (2,2) , (1,4) \\ 
1_1 \hookleftarrow (0,2) , (3,2) &\qquad&  
2_1 \hookleftarrow (1,0) , (2,1) , (1,3) , (3,2) \\ 
1_2 \hookleftarrow (1,2) , (5,0) &\qquad&  
2_2 \hookleftarrow (0,1) , (1,2) , (3,1) , (2,3) \\ 
1_3 \hookleftarrow (3,0) , (0,3) &\qquad&  
2_3 \hookleftarrow (1,1) , (0,3) , (2,2) , (4,1) \\ 
1_4 \hookleftarrow (2,1) , (0,5) &\qquad&  
2_4 \hookleftarrow (0,2) , (2,1) , (4,0) , (1,3) \\ 
1_5 \hookleftarrow (2,0) , (2,3) &\qquad&  
2_5 \hookleftarrow (2,0) , (1,2) , (3,1) , (0,4)  
\end{array} 
$$ 
The modular operator value $\hat{T}$ on the vertices $\lambda = (\lambda_1, \lambda_2) \in \mathcal{A}_5$ is given in Table \ref{T_A5}. 
\begin{table}[hhh] 
\small 
$$ 
\begin{array}{|c||c|c|c|c|c|c|c|c|c|c|c|c|} 
\hline 
(\lambda_1,\lambda_2) & (0,0) & (1,0) &(2,0) &(3,0) &(4,0) &(5,0) 
&(1,1) &(2,1) &(3,1) &(4,1) &(2,2) &(3,2) \\ 
                        &  {}   & (0,1) &(0,2) &(0,3) &(0,4) &(0,5) & 
{}   &(1,2) &(1,3) &(1,4) & {}   &(2,3) \\ 
\hline 
\hline 
\hat{T}  & 5 & 1 & 19 & 11 & 1 & 13 & 20 & 13 & 4 & 17 & 5 & 19 \\ 
\hline 
\end{array} 
$$ 
\caption{Value of $\hat{T}$ on the vertices $(\lambda_1, \lambda_2)$ of the $\mathcal{A}_5$ graph.} 
\normalsize 
\label{T_A5} 
\end{table} 
We see that the value of $\hat{T}$ on the vertices $\lambda = (\lambda_1,\lambda_2)$ whose restriction to $\mathcal{E}_5$ gives a vertex $1_i$ are equal. This allows us to define a fixed value of $\hat{T}$ from this induction mechanism to the subset $\mathcal{B}$ generated by the vertices 
$1_i$. This is not the case for its complement $\overline{\mathcal{B}}$ of $\mathcal{E}_5$. The subset 
$\mathcal{B}$ is also a subalgebra of the graph algebra of $\mathcal{E}_5$. We have: $\mathcal{E}_5 = 
\mathcal{B} \oplus \overline{\mathcal{B}}, \mathcal{B} \cdot \mathcal{B} \subset \mathcal{B}, \mathcal{B} \cdot \overline{\mathcal{B}} \subset \overline{\mathcal{B}}$.  
The modular invariant $\mathcal{M}$ is given by: 
\begin{equation} 
\mathcal{M}_{\lambda\mu} = \sum_{c \in \mathcal{B}} (F_{\lambda})_{1_0 c}(F_{\lambda})_{1_0 c } \;.
\end{equation} 
The extended characters $\hat{\chi}_a^5$, for $a \in \mathcal{B}$, are: 
\begin{equation} 
\hat{\chi}_a^5 = \sum_{\lambda} (F_{\lambda})_{1_0 a}\; \chi_{\lambda}^5 = \sum_{\lambda} 
(E_{1_0})_{\lambda a} \;\chi_{\lambda}^5 \;. 
\end{equation} 
The corresponding block-diagonal modular invariant partition function is: 
\begin{eqnarray*} 
\mathcal{Z}({\mathcal{E}_5}) =  
\sum_{c \in \mathcal{B}} |\hat{\chi}_{c}^5|^2 &=& |\chi_{(0,0)}^5 + \chi_{(2,2)}^5|^2 + |\chi_{(0,2)}^5 + \chi_{(3,2)}^5|^2 
 + |\chi_{(2,0)}^5 + \chi_{(2,3)}^5|^2 \\ 
{ } &+& |\chi_{(2,1)}^5 + \chi_{(0,5)}^5|^2 + |\chi_{(3,0)}^5 + \chi_{(0,3)}^5|^2 + |\chi_{(1,2)}^5 + \chi_{(5,0)}^5|^2 \; .
\end{eqnarray*} 

\paragraph{Conjugation and twist}
From the conjugation defined on the vertices of the $\mathcal{A}_5$ graph: $C(\lambda_1,\lambda_2)=(\lambda_2,\lambda_1)$, we can define the conjugation on vertices of the $\mathcal{E}_5$ graph, such that it is compatible with the induction-restriction mechanism between $\mathcal{A}_5$ and $\mathcal{E}_5$. For example, we have $1_0 \hookleftarrow (0,0) , (2,2)$. The vertices
(0,0) and (2,2) being self-conjugate, we set $1_0^*=1_0$. In the same way, 
$1_1 \hookleftarrow (0,2) , (3,2)$, and $1_5 \hookleftarrow (2,0) , (2,3)$, so we set
$1_1^*=1_5$. The conjugation on the $\mathcal{E}_5$ graph corresponds to the symmetry axis going through vertices $1_0$ and $1_3$ (the only real irreps): $1_{0}^{*}=1_0$,\mbox{$1_1^*=1_5$},\mbox{$1_2^*=1_4$},\mbox{$1_3^*=1_3$} ; $2_0^*=2_3,2_1^*=2_2,2_4^*=2_5$. As it should, 
conjugation corresponds matricially to transposition $G_{a^*} = G_a^{tr}$. 
The subset of vertices of $\mathcal{E}_5$ with well-defined modular operator value is 
$\mathcal{B} =\{1_i\}, i=0,5$. The conjugation is a $\hat{T}$-invariant involution: 
$\hat{T}(1_i) = \hat{T}(1_i^*)$. The modular invariant of the conjugated 
$\mathcal{E}_5$ case is given by:
\begin{equation} 
\mathcal{M}_{\lambda\mu} =  \sum_{c \in \mathcal{B}} (F_{\lambda})_{1_0 c}(F_{\mu})_{1_0 c^* } \;,
\end{equation} 
and the corresponding modular invariant partition function reads:
\begin{eqnarray*} 
\mathcal{Z}(\mathcal{E}_5^*) =  
\sum_{c \in \mathcal{B}} \hat{\chi}_{c}^5 \, \overline{\hat{\chi}_{c^*}^5} &=& |\chi_{(0,0)}^5 + 
\chi_{(2,2)}^5|^2 + 
|\chi_{(3,0)}^5 + \chi_{(0,3)}^5|^2  + (\chi_{(0,2)}^5 + \chi_{(3,2)}^5)
(\overline{\chi_{(2,0)}^5}+\overline{\chi_{(2,3)}^5}) 
\\ 
{ } &+& 
(\chi_{(2,0)}^5 + \chi_{(2,3)}^5)(\overline{\chi_{(0,2)}^5}+\overline{\chi_{(3,2)}^5}) + 
(\chi_{(1,2)}^5 + \chi_{(5,0)}^5)(\overline{\chi_{(0,5)}^5}+\overline{\chi_{(2,1)}^5}) \\ 
{ } &+&  (\chi_{(2,1)}^5 + \chi_{(0,5)}^5)(\overline{\chi_{(1,2)}^5}+
\overline{\chi_{(5,0)}^5}) \; .
\end{eqnarray*} 
It is associated to the conjugated graph $\mathcal{E}_5^*$ of reference \cite{DiF_Zub-graphs, Oc-Bariloche}, which is a module graph.  \\

The twist operator $\rho$ is defined on the $\mathcal{A}_5$ graph by $\rho(\lambda)=A^{5t(\lambda)}$, where $A$ is the $\mathbb{Z}_3$-symmetry of $\mathcal{A}_5$and $t(\lambda)$ is the triality. We can also define the operator $\rho$ on vertices of the $\mathcal{E}_5$ graph, in such a way that it is compatible with the induction-restriction mechanism between $\mathcal{A}_5$ and 
$\mathcal{E}_5$. Notice that the $\mathcal{E}_5$ graph possesses itself a $\mathbb{Z}_3$ symmetry, that corresponds to a rotation of angle $\pi/3$ around its center. The action of the $\mathbb{Z}_3$ symmetry on its vertices, that we still denote by $A$, is given by: 
$A(1_i)=1_{i+2} \mod6$, $A(2_i)=2_{i+2} \mod6$. We find that the twist operator $\rho$ is defined on the $\mathcal{E}_5$ graph in the same way as for $\mathcal{A}_5$: 
$\rho(a)=A^{5t(a)}(a)$, where $t(a)$ is the triality of the vertex $a \in \mathcal{E}_5$.
The twist operator $\rho$ is an involution and is $\hat{T}$-invariant for the vertices belonging to $\mathcal{B}$ (those with well-defined $\hat{T}$). The partition function defined by 
$\mathcal{Z}=\sum_{c \in \mathcal{B}} \hat{\chi}_c^5 \overline{\hat{\chi}_{\rho(c)}^5}$
is modular invariant. The graph associated, an orbifold of the $\mathcal{E}_5$ graph,
is denoted by $\mathcal{E}_5/3$. Notice that we have $\rho(c) = c^*$ for $c \in \mathcal{B}$, therefore the twist operator and the conjugation are the same on the subset $\mathcal{B}$, and  the corresponding partition functions are equal: $\mathcal{Z}(\mathcal{E}_5^*)=\mathcal{Z}(\mathcal{E}_5/3)$.
In fact, the module graphs $\mathcal{E}_5^*$ and $\mathcal{E}_5/3$ are the same: 
$\mathcal{E}_5^* = \mathcal{E}_5/3$.  We could also combine conjugation and the twist operator, but this doest not produce a new modular invariant.


\subsection{The $\mathcal{E}_9$ graph} 

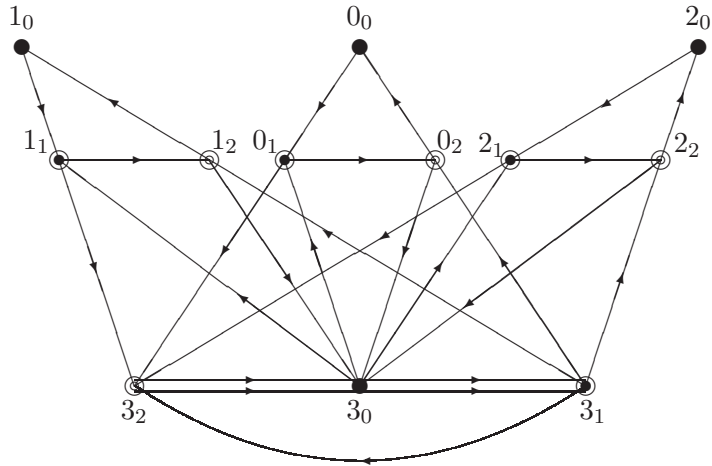
\begin{figure}[hhh] 
\unitlength=0.5mm 
\begin{center} 
 
\begin{picture}(180,130)(0,-20) 
\put(70,60){\begin{picture}(40,30) 
\put(0,0){\circle*{3}} 
\put(0,0){\circle{5}} 
\put(40,0){\circle{5}}  \put(40,0){\circle{2}} 
\put(20,30){\circle*{4}} 
\put(0,0){\vector(1,0){22.5}} 
\put(20,0){\line(1,0){20}} 
\put(40,0){\vector(-2,3){11.5}} 
\put(30,15){\line(-2,3){10}} 
\put(20,30){\vector(-2,-3){11.5}} 
\put(10,15){\line(-2,-3){10}} 
\end{picture}} 
 
\qbezier[920](30,0)(90,-40)(150,0) 
\put(90,-20){\vector(-1,0){0}} 
 
\put(30,0){\circle{5}} \put(30,0){\circle{2}} 
\put(90,0){\circle*{4}} 
\put(150,0){\circle*{3}} 
\put(150,0){\circle{5}} 
\put(10,60){\circle*{3}} 
\put(10,60){\circle{5}} 
\put(50,60){\circle{5}} \put(50,60){\circle{2}} 
\put(130,60){\circle*{3}} 
\put(130,60){\circle{5}} 
\put(170,60){\circle{5}}  \put(170,60){\circle{2}} 
\put(0,90){\circle*{4}} 
\put(180,90){\circle*{4}}

\put(0,90){\vector(1,-3){6.0}} 
\put(10,60){\line(-1,3){4.5}} 
\put(10,60){\vector(1,0){22.5}} 
\put(50,60){\line(-1,0){17.5}} 
 
\put(0,90){\line(5,-3){150}} 
\put(23,76.4){\vector(-3,2){0}} 
 
\put(30,0){\line(-1,3){10.8}} 
\put(10,60){\vector(1,-3){10}} 
 
\put(110,60){\vector(-1,-3){8.6}} 
\put(90,0){\line(1,3){15}} 
\put(90,0){\vector(-1,3){13}} 
\put(70,60){\line(1,-3){10}} 
 
\put(170,60){\line(-1,-3){10.8}} 
\put(150,0){\vector(1,3){10}} 
 
\put(180,90){\line(-5,-3){150}} 
\put(154,74){\vector(-3,-2){0}} 
 
\put(130,60){\vector(1,0){22.5}} 
\put(170,60){\line(-1,0){17.5}} 
\put(170,60){\vector(1,3){6.0}} 
\put(180,90){\line(-1,-3){4.5}} 
 
\put(70,60){\line(-2,-3){40}} 
\put(70,60){\vector(-2,-3){17}} 
\put(110,60){\line(2,-3){40}} 
\put(150,0){\vector(-2,3){23}} 
 
\put(30,1.5){\line(1,0){120}} 
\put(30,-1.5){\line(1,0){120}} 
\put(30,1.5){\vector(1,0){32.5}} 
\put(30,-1.5){\vector(1,0){32.5}} 
\put(90,1.5){\vector(1,0){32.5}} 
\put(90,-1.5){\vector(1,0){32.5}} 
 
\put(90,0){\line(-4,3){80}} 
\put(90,0){\vector(-4,3){32.5}} 
\put(90,0){\line(4,3){80}} 
\put(170,60){\vector(-4,-3){52.5}} 
 
\put(50,60){\line(2,-3){40}} 
\put(50,60){\vector(2,-3){22}} 
\put(90,0){\line(2,3){40}} 
\put(90,0){\vector(2,3){22}} 
 
\put(80,41.7){\vector(-3,2){0}} 
\put(95,38.7){\vector(-3,-2){0}} 
 
\put(0,98){\makebox(0,0){$1_0$}} 
\put(90,98){\makebox(0,0){$0_0$}} 
\put(180,98){\makebox(0,0){$2_0$}} 
 
\put(4,65){\makebox(0,0){$1_1$}} 
\put(65,65){\makebox(0,0){$0_1$}} 
\put(125,64){\makebox(0,0){$2_1$}} 
 
\put(54,65){\makebox(0,0){$1_2$}} 
\put(114,65){\makebox(0,0){$0_2$}} 
\put(177,64){\makebox(0,0){$2_2$}} 
 
\put(30,-7){\makebox(0,0){$3_2$}} 
\put(90,-7){\makebox(0,0){$3_0$}} 
\put(152,-7){\makebox(0,0){$3_1$}}

\end{picture} 
\end{center} 
\caption{The $\mathcal{E}_9$ graph.} 
\label{fig:graphE9}
\end{figure}

The $\mathcal{E}_{9}$ graph has $12$ vertices denoted by 
$0_{i},1_{i},2_{i},3_{i}$ with $i=0,1,2$. 
Its level $k=9$ and altitude $\kappa = 12$. The identity is $0_{0}$ and the left and right
chiral generators are $0_{1}$ and $0_{2}$. The graph is 3--colorable, the indice $i$ giving the triality of the vertex $a_i$. The graph is displayed in Figure \ref{fig:graphE9} with the orientation corresponding to the generator $0_1$.

The graph of the $\mathcal{A}$ series with same norm is
$\mathcal{A}_9$ with $(k+1)(k+2) /2=55$ vertices labelled by $\lambda= (\lambda_1,\lambda_2)$.
The Coxeter exponents of $\mathcal{E}_{9}$ are given by:
\begin{equation}\label{CoexpE9}
(r_{1},r_{2}) = (0,0),\; (4,4),\; (1,4),\; (4,1),\; (9,0),\; (0,9),\;  
\textrm{ and twice } \; (2,2),\; (2,5),\; (5,2),\; 
\end{equation}

The $\mathcal{E}_9$ graph is a subgroup graph. From the graph we can deduce its graph algebra structure and its graph algebra matrices $G_a$. Notice that we have to impose that the structure coefficients are positive integers in order to get a unique result (see discussion in \cite{Coq_Gil-Tmod}). The fused 
matrices $F_{\lambda}$ are calculated from the truncated SU(3) recursion formula with starting point $F_{(1,0)}=G_{0_1}$. The essential matrix $E_{0_0}$ has 55 lines labelled by vertices
$\lambda \in \mathcal{A}_{9}$ and 12 columns labelled by vertices of $\mathcal{E}_9$. 
From this essential matrix we read the induction 
$\mathcal{E}_9 \hookleftarrow \mathcal{A}_9$. For the vertices $0_0, 1_0$ and $2_0$ it is given by:
\begin{eqnarray*} 
0_0 &\hookleftarrow  & (0,0) , (4,4) , (4,1) , (1,4) , (9,0) , (0,9) \\ 
1_0 &\hookleftarrow  & (2,2) , (5,2) , (2,5) \\ 
2_0 &\hookleftarrow  & (2,2) , (5,2) , (2,5)  
\end{eqnarray*} 
We can assign a fixed modular operator value to these vertices: 
$\hat{T}(0_0) = 9$ and $\hat{T}(1_0)=\hat{T}(2_0)=21$. But this is not the case for the other vertices. The subset $\mathcal{B}$ is defined in by $\mathcal{B} = \{0_0,1_0,2_0\}$. It is a subalgebra of the graph algebra of $\mathcal{E}_9$.  
The modular invariant of $\mathcal{E}_9$ is defined by:
\begin{equation}
\mathcal{M}_{\lambda\mu} = \sum_{c \in \mathcal{B}} (F_{\lambda})_{0_0 c} (F_{\mu})_{0_0 c} \;.
\end{equation}
The extended characters $\hat{\chi}_c^9$ are given by:
\begin{eqnarray*} 
\hat{\chi}_{0_0}^9  &=& \chi_{0,0}^9 + \chi_{0,9}^9 + \chi_{9,0}^9 + \chi_{1,4}^9 + 
\chi_{4,1}^9 + \chi_{4,4}^9 \;, \\ 
\hat{\chi}_{1_0}^9 = \hat{\chi}_{2_0}^9 &=& \chi_{2,2}^9 + \chi_{2,5}^9 + \chi_{5,2}^9   \;,
\end{eqnarray*} 
and the modular invariant partition function reads:
\begin{equation}
\mathcal{Z}(\mathcal{E}_9) = \sum_{c \in \mathcal{B}} |\hat{\chi}_c^9|^2 
= |\hat{\chi}_{0_0}^9|^2 + 2 \, |\hat{\chi}_{1_0}^9|^2 \;.
\end{equation}

\paragraph{Conjugation and twist} From the conjugation and the twist operator $\rho$ defined on the $\mathcal{A}_9$ graph, we can define their analogue on the $\mathcal{E}_9$ graph, in the same way as for the $\mathcal{E}_5$ graph. Conjugation is well-defined: vertices of triality 0 are self-conjugated $a_0^*=a_0$, and $a_1^* = a_2$. The twist operator defined from the 
$\mathcal{A}_9$ graph however is trivial, since here the level is $k=9$ and $A^3=\munite$. The conjugation $C$ is a $\hat{T}$-invariant involution for vertices 
$c \in \mathcal{B}$. Notice that we have $c^*=c$ for $c \in \mathcal{B}$, the conjugation reduces to the identity in $\mathcal{B}$, therefore we find $\mathcal{Z}(\mathcal{E}_9)=\mathcal{Z}(\mathcal{E}_9^*)$. However, the subgroup graph $\mathcal{E}_9$ and the module graph  $\mathcal{E}_9^*$ of references \cite{DiF_Zub-graphs, Oc-Bariloche} are different, event if they are associated to the same $\mathcal{Z}$. Notice that vertices $1_0$ and $2_0$ have the same modular operator value $\hat{T}$. We can therefore define a $\hat{T}$-invariant involution $\rho$ such that $\rho(0_0)=0_0$ and $\rho(1_0)=2_0$ on 
vertices $\in \mathcal{B}$ (this $\rho$ is not the one coming from the $\mathbb{Z}_3$-symmetry of $\mathcal{A}_9$). But notice that $\hat{\chi}_c^9 = \hat{\chi}_{\rho(c)}^9$, so the corresponding modular invariant partition function $\mathcal{Z}(\mathcal{E}_9^{\rho}) = 
\sum_{c \in \mathcal{B}} \hat{\chi}_c^9 \overline{\hat{\chi}_{\rho(c)}^9} = \mathcal{Z}(\mathcal{E}_9)$.
In fact, the $\mathcal{E}_9$ graph possesses a $\mathbb{Z}_3$-symmetry, corresponding to the symmetry around the axis formed by vertices $3_0,3_1$ and $3_2$ ($\mathbb{Z}_3$-invariant vertices).  If Figure \ref{fig:graphE9} was drawn in 3D, the $\mathbb{Z}_3$-symmetry 
will correspond to the symmetry around the axis formed by these vertices. The graph associated to $\mathcal{Z}(\mathcal{E}_9^{\rho})$ is the $\mathbb{Z}_3$-orbifold of the $\mathcal{E}_9$ graph.  These fixed vertices are triplicated in the orbifold graph,
and the other vertices are identified with their orbit. The orbifold graph (module graph) has
therefore $3+(3\times3)=12$ vertices and is denoted by $\mathcal{E}_9/3$ in \cite{Oc-Bariloche}.
It is equal to the conjugated graph $\mathcal{E}_9^*$. Its adjacency matrix can be calculated from the orbifold mechanism described in the $\mathcal{D}_k$ section.


\subsection{The ${\mathcal{E}}_{21}$ case}
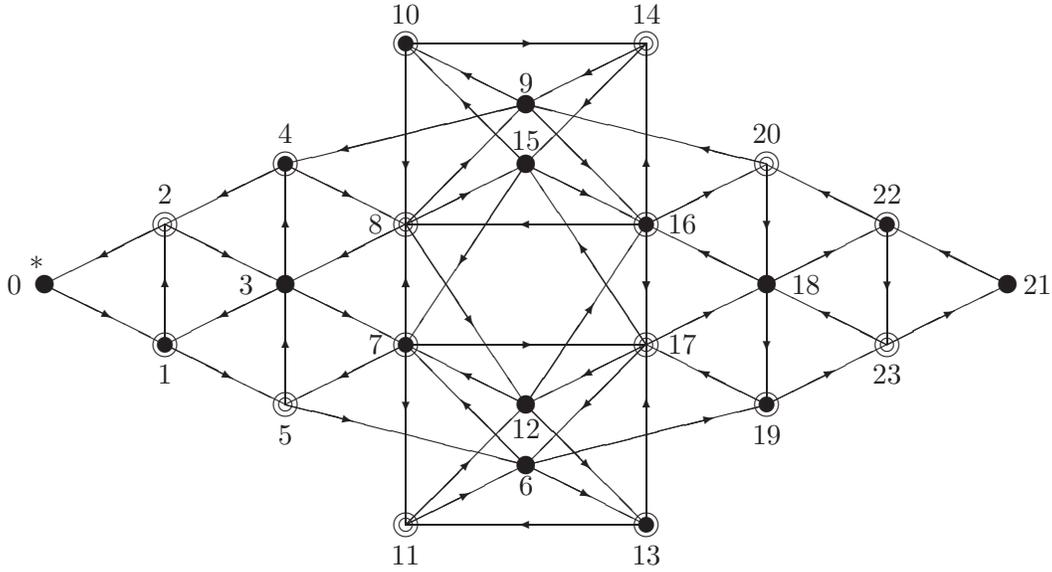
\begin{figure}[hhh] 
\unitlength 0.40mm 
\begin{center} 
\begin{picture}(320,170) 
 
\put(0,60){\begin{picture}(40,40) 
\put(0,20){\circle*{6}} 
\put(40,0){\circle*{5}} 
\put(40,0){\circle{8}} 
\put(40,40){\circle{8}} 
\put(40,40){\circle{4}} 
\put(0,20){\line(2,-1){40}} 
\put(0,20){\line(2,1){40}} 
\put(40,0){\line(0,1){40}} 
\put(40,40){\vector(-2,-1){22.5}} 
\put(0,20){\vector(2,-1){22.5}} 
\put(40,0){\vector(0,1){22.5}} 
\end{picture}} 
 
\put(40,40){\begin{picture}(40,40) 
\put(40,0){\circle{8}} 
\put(40,0){\circle{4}} 
\put(40,40){\circle*{6}} 
\put(0,20){\line(2,-1){40}} 
\put(0,20){\line(2,1){40}} 
\put(40,0){\line(0,1){40}} 
\put(40,40){\vector(-2,-1){22.5}} 
\put(0,20){\vector(2,-1){22.5}} 
\put(40,0){\vector(0,1){22.5}} 
\end{picture}} 
 
\put(40,80){\begin{picture}(40,40) 
\put(40,40){\circle*{5}} 
\put(40,40){\circle{8}} 
\put(0,20){\line(2,-1){40}} 
\put(0,20){\line(2,1){40}} 
\put(40,0){\line(0,1){40}} 
\put(40,40){\vector(-2,-1){22.5}} 
\put(0,20){\vector(2,-1){22.5}} 
\put(40,0){\vector(0,1){22.5}} 
\end{picture}} 
 
\put(80,60){\begin{picture}(40,40) 
\put(40,0){\circle*{5}} 
\put(40,0){\circle{8}} 
\put(40,40){\circle{8}} 
\put(40,40){\circle{4}} 
\put(0,20){\line(2,-1){40}} 
\put(0,20){\line(2,1){40}} 
\put(40,0){\line(0,1){40}} 
\put(40,40){\vector(-2,-1){22.5}} 
\put(0,20){\vector(2,-1){22.5}} 
\put(40,0){\vector(0,1){22.5}} 
\end{picture}} 
 
\put(80,40){\line(4,-1){80}} 
\put(80,40){\vector(4,-1){22.5}} 
\put(80,120){\line(4,1){80}} 
\put(160,140){\vector(-4,-1){62.5}} 

\put(280,60){\begin{picture}(40,40) 
\put(0,0){\circle{8}} 
\put(0,0){\circle{4}} 
\put(40,20){\circle*{6}} 
\put(0,40){\circle*{5}} 
\put(0,40){\circle{8}} 
\put(0,0){\line(0,1){40}} 
\put(0,0){\line(2,1){40}} 
\put(0,40){\line(2,-1){40}} 
\put(0,0){\vector(2,1){22.5}} 
\put(40,20){\vector(-2,1){22.5}} 
\put(0,40){\vector(0,-1){22.5}} 
\end{picture}} 
 
\put(240,40){\begin{picture}(40,40) 
\put(0,0){\circle*{5}} 
\put(0,0){\circle{8}} 
\put(0,40){\circle*{6}} 
\put(0,0){\line(0,1){40}} 
\put(0,0){\line(2,1){40}} 
\put(0,40){\line(2,-1){40}} 
\put(0,0){\vector(2,1){22.5}} 
\put(40,20){\vector(-2,1){22.5}} 
\put(0,40){\vector(0,-1){22.5}} 
\end{picture}} 
 
\put(240,80){\begin{picture}(40,40) 
\put(0,40){\circle{8}} 
\put(0,40){\circle{4}} 
\put(0,0){\line(0,1){40}} 
\put(0,0){\line(2,1){40}} 
\put(0,40){\line(2,-1){40}} 
\put(0,0){\vector(2,1){22.5}} 
\put(40,20){\vector(-2,1){22.5}} 
\put(0,40){\vector(0,-1){22.5}} 
\end{picture}} 
 
\put(200,60){\begin{picture}(40,40) 
\put(0,0){\circle{8}} 
\put(0,0){\circle{4}} 
 
\put(0,40){\circle*{5}} 
\put(0,40){\circle{8}} 
\put(0,0){\line(0,1){40}} 
\put(0,0){\line(2,1){40}} 
\put(0,40){\line(2,-1){40}} 
\put(0,0){\vector(2,1){22.5}} 
\put(40,20){\vector(-2,1){22.5}} 
\put(0,40){\vector(0,-1){22.5}} 
\end{picture}} 
\put(160,20){\line(4,1){80}} 
\put(160,20){\vector(4,1){62.5}} 
 
\put(240,120){\line(-4,1){80}} 
\put(240,120){\vector(-4,1){22.5}}

\put(120,0){\circle{8}} 
\put(120,0){\circle{4}} 
\put(200,0){\circle*{5}} 
\put(200,0){\circle{8}} 
\put(160,20){\circle*{6}}

\put(160,40){\circle*{6}} 
\put(160,120){\circle*{6}} 
\put(160,140){\circle*{6}} 
\put(120,160){\circle*{5}} 
\put(120,160){\circle{8}} 
\put(200,160){\circle{8}} 
\put(200,160){\circle{4}}

\put(200,0){\line(-1,0){80}} 
\put(200,0){\vector(-1,0){42.5}} 
\put(120,0){\line(2,1){40}} 
\put(120,0){\line(1,1){40}} 
\put(120,0){\vector(2,1){21.5}} 
\put(120,0){\vector(1,1){21.5}} 
\put(160,20){\line(2,-1){40}} 
\put(160,20){\vector(2,-1){21.5}} 
\put(160,40){\line(1,-1){40}} 
\put(160,40){\vector(1,-1){21.5}} 
\put(200,100){\line(-1,0){80}} 
\put(200,100){\vector(-1,0){42.5}} 
\put(120,100){\line(2,1){40}} 
\put(120,100){\line(1,1){40}} 
 
\put(120,100){\vector(2,1){21.5}} 
\put(120,100){\vector(1,1){21.5}} 
\put(160,120){\line(2,-1){40}} 
\put(160,120){\vector(2,-1){21.5}} 
\put(160,140){\line(1,-1){40}} 
\put(160,140){\vector(1,-1){21.5}}

\put(120,160){\line(1,0){80}} 
\put(120,160){\vector(1,0){42.5}} 
\put(200,160){\line(-2,-1){40}} 
\put(200,160){\line(-1,-1){40}} 
\put(200,160){\vector(-2,-1){21.5}} 
\put(200,160){\vector(-1,-1){21.5}} 
\put(160,140){\line(-2,1){40}} 
\put(160,140){\vector(-2,1){21.5}} 
\put(160,120){\line(-1,1){40}} 
\put(160,120){\vector(-1,1){21.5}} 
 
\put(120,60){\line(1,0){80}} 
\put(120,60){\vector(1,0){42.5}} 
\put(200,60){\line(-2,-1){40}} 
\put(200,60){\line(-1,-1){40}} 
\put(200,60){\vector(-2,-1){21.5}} 
\put(200,60){\vector(-1,-1){21.5}} 
\put(160,40){\line(-2,1){40}} 
\put(160,40){\vector(-2,1){21.5}} 
\put(160,20){\line(-1,1){40}} 
\put(160,20){\vector(-1,1){21.5}}

\put(120,0){\line(0,1){160}} 
\put(200,0){\line(0,1){160}} 
\put(120,60){\vector(0,1){22.5}} 
\put(120,60){\vector(0,-1){22.5}} 
\put(120,160){\vector(0,-1){42.5}} 
\put(200,100){\vector(0,-1){22.5}} 
\put(200,100){\vector(0,1){22.5}} 
\put(200,0){\vector(0,1){42.5}}

\put(200,60){\line(-2,3){40}} 
\put(200,60){\vector(-2,3){22.5}} 
\put(160,120){\line(-2,-3){40}} 
\put(160,120){\vector(-2,-3){22.5}}

\put(120,100){\line(2,-3){40}} 
\put(120,100){\vector(2,-3){22.5}} 
\put(160,40){\line(2,3){40}} 
\put(160,40){\vector(2,3){22.5}} 
 
\put(120,60){\line(-2,-1){40}} 
\put(120,60){\vector(-2,-1){22.5}} 
\put(80,120){\line(2,-1){40}} 
\put(80,120){\vector(2,-1){22.5}} 
 
\put(240,40){\line(-2,1){40}} 
\put(240,40){\vector(-2,1){22.5}} 
\put(200,100){\line(2,1){40}} 
\put(200,100){\vector(2,1){22.5}}

\put(-10,80){\makebox(0,0){0}} 
\put(67,80){\makebox(0,0){3}} 
\put(253,80){\makebox(0,0){18}} 
\put(330,80){\makebox(0,0){21}} 
 
\put(40,50){\makebox(0,0){1}} 
\put(40,110){\makebox(0,0){2}} 
 
\put(80,30){\makebox(0,0){5}} 
\put(80,130){\makebox(0,0){4}} 
\put(120,-10){\makebox(0,0){11}} 
\put(120,170){\makebox(0,0){10}} 
\put(200,-10){\makebox(0,0){13}} 
\put(200,170){\makebox(0,0){14}} 
\put(240,30){\makebox(0,0){19}} 
\put(240,130){\makebox(0,0){20}} 
\put(280,50){\makebox(0,0){23}} 
\put(280,110){\makebox(0,0){22}} 
\put(160,13){\makebox(0,0){6}} 
\put(160,32){\makebox(0,0){12}} 
\put(160,128){\makebox(0,0){15}} 
\put(160,147){\makebox(0,0){9}} 
\put(110,60){\makebox(0,0){7}} 
\put(110,100){\makebox(0,0){8}} 
\put(212,60){\makebox(0,0){17}} 
\put(212,100){\makebox(0,0){16}}

\put(-5,85){$\ast$} 
 
\end{picture} 
\end{center} 
\caption{The $\mathcal{E}_{21}$ graph.} 
\label{fig:graphE21}
\end{figure}

The $\mathcal{E}_{21}$ graph has $24$ vertices
labelled by $a \in \{0,1,2,\cdots,23\}$, 
its level $k=21$, altitude $\kappa=24$ and its norm $\beta = 1+ 2\cos(2 \pi / 24) = 
1+\frac{1}{2 }\sqrt{2}+\frac{1}{2}\sqrt{6}$.
$\mathcal{E}_{21}$ graph is $3-$ colorable, the triality of a vertex $a$ is $t(a)=a$ mod 3. 
The unity is $0$ and the left and right chiral generators are $1$ and $2$. The graph is 
displayed in Figure \ref{fig:graphE21} with the orientation corresponding to the 
left generator 1.

The graph of the $\mathcal{A}$ series with same norm is $\mathcal{A}_{21}$,
with $22 \times 23 /2 = 253$ vertices.
The eigenvalues of the adjacency matrix of $\mathcal{E}_{21}$ is a subset of those of the 
$\mathcal{A}_{21}$ graph. The Coxeter exponents of $\mathcal{E}_{21}$ are:
\begin{equation}\label{CoexpE21}
\begin{array}{rcl}
(r_{1},r_{2}) &=& (0,0) \; , \; (21,0) \; , \; (0,21) \; , \; (4,4) \; , \; (13,4) \; , \; (4,13) \; , \; (6,6) \; , \; (9,6) \\
{ } &{ }& (6,9) \; , \; (10,10) \; , \; (10,1) \; , \; (1,10) \; , \; (6,0) \; , \; (15,6) \; , \; (0,15) \; , \; (0,6) \\
{ } &{ }& (15,0) \; , \; (6,15) \; , \; (4,7) \; , \; (10,4) \; , \; (7,10) \; , \; (7,4) \; , \; (10,7) \; , \; (4,10) \;.
\end{array}
\end{equation}

The $\mathcal{E}_{21}$ graph is a subgroup graph, it determines in a unique way its graph 
algebra. The determination of the graph algebra matrices $G_a$ is straightforward up to vertex 9 (for example
$1 \cdot 1=2+5$ gives $G_5=G_1.G_1-G_2$). The graph possesses a symmetry around the center of the
central star. We call $a'$ the symmetric of $a$ (for example 0'=21, 1'=22). At multiplication level, we have the following property:
\begin{equation}
a\cdot b=a'\cdot b'
\end{equation}
Multiplication of a vertex $a$ by the vertex $21$ gives its symmetric $a'$, i.e 
$G_{a'} = G_{21} . G_{a}$. Such data allows us to compute easily all the graph
algebra matrices. $G_0$ is the identity matrix and $G_1$ is the adjacency matrix of the graph. The others are given by the following polynomials:
\begin{equation}
\begin{array}{l}
G_{2}=(G_{1})^{tr} \, , \; 
G_{3}=G_{1}(G_{1})^{tr}-G_{0} \, , \; 
(G_{4})^{tr}=G_{5}=G_{1}G_{1}-G_{2} \, , \;
(G_{9})^{tr}=G_{6}=G_{1}G_{5}-G_{3} \, , 
\\
(G_{8})^{tr}=G_{7}=G_{2}G_{5}-G_{1} \, , \;   
G_{21}=G_{9}G_{6}^{-1} \, , \; G_{18}=G_{21}G_{3} \, , \;
(G_{17})^{tr}=G_{16}=G_{21}G_{7} \, ,
\\
(G_{19})^{tr}=G_{20}=G_{21}G_{5} \, , \; 
(G_{23})^{tr}=G_{22}=G_{21}G_{1} \, , \;
(G_{10})^{tr}=G_{11}=G_{1}G_{7}-G_{5}-G_{8}-G_{17} \, ,
\\ 
(G_{15})^{tr}=G_{12}=G_{1}G_{11}-G_{6} \, , \; 
(G_{14})^{tr}=G_{13}=G_{21}G_{10} \;.
\end{array}
\end{equation}
 
\paragraph{Induction-restriction} 
The action of the graph algebra $\mathcal{A}_{21}$ on $\mathcal{E}_{21}$ is coded by the 253 matrices $F_{\lambda} = F_{(\lambda_1, \lambda_2)}$ obtained by the truncated SU(3) recursion formulae, with $F_{(1,0)} = G_1$. The essential matrix (intertwiner) $E_0$ has 253 lines 
labelled by vertices of $\mathcal{A}_{21}$ and 24 columns labelled by vertices of 
$\mathcal{E}_{21}$, it gives the induction-restriction rules between $\mathcal{A}_{21}$ and $\mathcal{E}_{21}$. We can verify that the values of the modular operator $\hat{T}$ is constant  over the vertices of $\mathcal{A}_{21}$ whose restriction to $\mathcal{E}_{21}$ gives the vertex 0 and 21. We find $\hat{T}(0)=3$ and $\hat{T}(21)=57$. This is not the case for the other vertices. This induction mechanism characterizes the subset $\mathcal{B} = \{0,21\}$. This subset is a subalgebra of $\mathcal{E}_{21}$. 
The modular invariant $\mathcal{M}$ is given by: 
\begin{equation} 
\mathcal{M}_{\lambda\mu} = \sum_{c \in \mathcal{B}} (F_{\lambda})_{0c} \, (F_{\mu})_{0c} \;,
\end{equation}
 and the extended characters $\hat{\chi_c}^{21}$, for $c \in \mathcal{B}$, by:
\footnotesize 
\begin{eqnarray*} 
\hat{\chi}_0^{21}   &=& \chi_{(0,0)}^{21} +  \chi_{(4,4)}^{21} + \chi_{(6,6)}^{21} + \chi_{(10,10)}^{21} + \chi_{(0,21)}^{21} + \chi_{(21,0)}^{21} +  \chi_{(1,10)}^{21} + \chi_{(10,1)}^{21} + \chi_{(4,13)}^{21} + \chi_{(13,4)}^{21} +  \chi_{(6,9)}^{21} + \chi_{(9,6)}^{21} \;, \\ 
\hat{\chi}_{21}^{21}&=& \chi_{(0,6)}^{21} +  \chi_{(6,0)}^{21} + \chi_{(0,15)}^{21} + \chi_{(15,0)}^{21} + \chi_{(4,7)}^{21} + \chi_{(7,4)}^{21} +   
\chi_{(4,10)}^{21} + \chi_{(10,4)}^{21} + \chi_{(6,15)}^{21} + \chi_{(15,6)}^{21} + \chi_{(7,10)}^{21} + \chi_{(10,7)}^{21} \,.
\end{eqnarray*} 
\normalsize 
The corresponding modular invariant partition function is given by: 
\begin{equation} 
\mathcal{Z}(\mathcal{E}_{21}) = \sum_{c \in \mathcal{B}} |\hat{\chi}_c^{21}|^2 = |\hat{\chi}_0^{21}|^2 + |\hat{\chi}_{21}^{21}|^2 \;.
\label{modinvZE21}
\end{equation}

\paragraph{Conjugation and twist}
Conjugation is well-defined on the $\mathcal{E}_{21}$ graph. It corresponds to the symmetry axis 
going through vertices $ \{0-3-18-21\}$. Vertices $\in \mathcal{B}$ are self-conjugated, 
so $\mathcal{Z}(\mathcal{E}_{21}^*) = \mathcal{Z}(\mathcal{E}_{21})$. There is no twist operator related to a $\mathbb{Z}_3$-symmetry. The one coming from the $\mathcal{A}_{21}$ graph is trivial since the level $k=21$ is modulo 3, and, unlike the $\mathcal{E}_9$ graph, the graph itself does not possess a $\mathbb{Z}_3$-symmetry. Furthemore, we have $\hat{T}(0)\not= \hat{T}(21)$, so no $\hat{T}$-invariant twist exists (except the identity). The only exceptional modular invariant partition function at this level is given by Equation (\ref{modinvZE21}).

\section{Conclusions and outlook}
The affine $su(3)$ modular invariants are associated to a set of generalized Coxeter graphs. Subgroup and module graphs, from their spectral properties, are related to Type I and Type II modular invariants respectively. But all modular invariants can be determined from the subset of subgroup graphs and suitable twists. We determine the ambichiral algebra $\mathcal{B}$ from induction-restriction maps and from modular properties of the subgroup graphs, defined through a modular operator $\hat{T}$ taking values on its set of vertices. We obtain all Type I modular invariants. The twists $\vartheta$ are the involutions of elements of $\mathcal{B}$ that leave invariant the modular operator $\hat{T}$. 
A simple analysis of all such twists allow us to obtain all type II modular invariants. 
 Thinking of generalizations towards higher rank affine algebras, we believe that the use of the modular operator $\hat{T}$ can simplify the association between graphs and modular invariants. It is also a useful tool for the construction of the quantum symmetries of the generalized Coxeter graphs.
\vspace{1cm}

\noindent {\bf \Large Acknowledgements}
D.H. and E.H.T. would like to thank R. Coquereaux (R.C.) to introduce them to the subject of graphs belonging to Coxeter-Dynkin systems and the study of their quantum symmetries. The authors would also like to thank R.C. for a carefull reading of the manuscript.  G.S. would like to thank FAPERJ -- Funda\c{c}\~ao de Amparo \`a Pesquisa do Estado do Rio de Janeiro -- for financial support.

\appendix

\section{The affine $su(3)$ modular invariants and the associated graphs}

\begin{table}[H]
\small
\begin{center}
$$
\begin{array}{|rcl|}
\hline
&& \\
\mathcal{A}_k  & & \mathcal{Z} = \dps \sum_{\lambda \in \mathcal{P}_+^k} |\chi_{\lambda}^k|^2   \\
\mathcal{A}_k^*  & & \mathcal{Z} = \dps \sum_{\lambda \in \mathcal{P}_+^k} \chi_{\lambda}^k
\,\overline{\chi_{\lambda}^k}    \\
\mathcal{D}_{k \equiv 1,2} & & \mathcal{Z} = \dps \sum_{\lambda \in \mathcal{P}_+^k}
\chi_{\lambda}^k \; \overline{\chi_{\rho(\lambda)}^{k}}  \\ 
\mathcal{D}_{k \equiv 1,2}^* & & \mathcal{Z} = \dps \sum_{\lambda \in \mathcal{P}_+^k}
\chi_{\lambda}^k \; \overline{\chi_{\rho(\lambda)^*}^{k}}  \\ 
\mathcal{D}_{k \equiv 0}  & & \mathcal{Z} = \dps \frac{1}{3} \sum_{\tiny \begin{array}{c} \lambda \in \mathcal{P}_+^k \\ t(\lambda)=0 \end{array}} |\chi_{\lambda}^k + \chi_{A(\lambda)}^k+\chi_{A^2(\lambda)}^k|^2   \\
\mathcal{D}_{k \equiv 0}^*  & & \mathcal{Z} = \dps \frac{1}{3} \sum_{\tiny \begin{array}{c} \lambda \in \mathcal{P}_+^k \\ t(\lambda)=0 \end{array}} (\chi_{\lambda}^k + \chi_{A(\lambda)}^k+\chi_{A^2(\lambda)}^k) \, (\overline{\chi_{\lambda^*}^k} + \overline{\chi_{A(\lambda)^*}^k} + \overline{\chi_{A^2(\lambda)^*}^k})  \\
\mathcal{D}_9^{\xi} & & \mathcal{Z} = |\chi_{(0,0)}^9+\chi_{(9,0)}^9+\chi_{(0,9)}^9|^2
+ |\chi_{(3,0)}^9+\chi_{(6,3)}^9+\chi_{(0,6)}^9|^2
+ |\chi_{(0,3)}^9+\chi_{(6,0)}^9+\chi_{(3,6)}^9|^2 \nonumber \\
 & & \qquad + |\chi_{(2,2)}^9+\chi_{(5,2)}^9+\chi_{(2,5)}^9|^2
+ |\chi_{(4,4)}^9+\chi_{(4,1)}^9+\chi_{(1,4)}^9|^2
+ 2\, |\chi_{(3,3)}^9|^2 \nonumber \\
&& \qquad + \left\lbrack(\chi_{(1,1)}^9+\chi_{(7,1)}^9+\chi_{(1,7)}^9)\, 
\overline{\chi_{(3,3)}^9} + h.c. \right\rbrack  \\
&& \\
\mathcal{D}_9^{\xi*} & & \mathcal{Z} = |\chi_{(0,0)}^9+\chi_{(9,0)}^9+\chi_{(0,9)}^9|^2
+ |\chi_{(2,2)}^9+\chi_{(5,2)}^9+\chi_{(2,5)}^9|^2
+ |\chi_{(4,4)}^9+\chi_{(4,1)}^9+\chi_{(1,4)}^9|^2 \nonumber \\
 & & \qquad   
+ 2\, |\chi_{(3,3)}^9|^2 + 
\left\lbrack (\chi_{(0,3)}^9+\chi_{(6,0)}^9 + \chi_{(3,6)}^9) \, 
(\overline{\chi_{(3,0)}^9} + \overline{\chi_{(6,3)}^9} + \overline{\chi_{(0,6)}^9}) + h.c.
\right\rbrack
+  \nonumber \\
&& \qquad + \left\lbrack (\chi_{(1,1)}^9+\chi_{(7,1)}^9+\chi_{(1,7)}^9)\, 
\overline{\chi_{(3,3)}^9} + h.c. \right\rbrack  \\
&& \\
\mathcal{E}_5 & & \mathcal{Z} = |\chi_{(0,0)}^5+\chi_{(2,2)}^5|^2 + 
|\chi_{(0,2)}^5+\chi_{(3,2)}^5|^2 + |\chi_{(2,0)}^5+\chi_{(2,3)}^5|^2 
+ |\chi_{(2,1)}^5+\chi_{(0,5)}^5|^2 \nonumber \\
{ }  & & \qquad  + |\chi_{(3,0)}^5+\chi_{(0,3)}^5|^2
+ |\chi_{(1,2)}^5+\chi_{(5,0)}^5|^2 \\
&& \\
\mathcal{E}_5^* = \mathcal{E}_5/3 & & \mathcal{Z} = |\chi_{(0,0)}^5+\chi_{(2,2)}^5|^2 + 
|\chi_{(3,0)}^5+\chi_{(0,3)}^5|^2 + 
\left\lbrack (\chi_{(0,2)}^5+\chi_{(3,2)}^5) \,  + (\overline{\chi_{(2,0)}^5}+\overline{\chi_{(2,3)}^5}) + h.c. \right\rbrack \nonumber \\
{ }  & & \qquad  + \left\lbrack (\chi_{(2,1)}^5+\chi_{(0,5)}^5) \, 
(\overline{\chi_{(1,2)}^5} + \overline{\chi_{(5,0)}^5}) + h.c. \right\rbrack \\
&& \\
\mathcal{E}_9 \, , \, \mathcal{E}_9^*=\mathcal{E}_9/3 & & \mathcal{Z} = |\chi_{(0,0)}^9 + \chi_{(0,9)}^9 + \chi_{(9,0)}^9 + 
\chi_{(1,4)}^9 + \chi_{(4,1)}^9 + \chi_{(4,4)}^9|^2 + 2 |\chi_{(2,2)}^9 + \chi_{(2,5)}^9 + \chi_{(5,2)}^9|^2 \qquad \quad \\
&& \\
\mathcal{E}_{21} & & \mathcal{Z} =  
|\chi_{(0,0)}^{21} + \chi_{(4,4)}^{21} + \chi_{(6,6)}^{21} + \chi_{(10,10)}^{21}+
 \chi_{(0,21)}^{21} + \chi_{(21,0)}^{21} +  \chi_{(1,10)}^{21} + \chi_{(10,1)}^{21} + \chi_{(4,13)}^{21}  \nonumber \\ 
{ } & & \qquad + \chi_{(13,4)}^{21}+
 \chi_{(6,9)}^{21} + \chi_{(9,6)}^{21}|^2 + |\chi_{(0,6)}^{21} + \chi_{(6,0)}^{21} + \chi_{(0,15)}^{21} + \chi_{(15,0)}^{21}+
 \chi_{(4,7)}^{21} + \chi_{(7,4)}^{21} \nonumber \\
{ } & & \qquad +  \chi_{(4,10)}^{21} + \chi_{(10,4)}^{21} + \chi_{(6,15)}^{21} + \chi_{(15,6)}^{21}+
 \chi_{(7,10)}^{21} + \chi_{(10,7)}^{21}|^2 \\
&& \\
\hline
\end{array}
$$
\end{center}
\caption{The affine $su(3)$ modular invariants and the associated Di Francesco-Zuber graphs. 
Subgroup graphs, related to block-diagonal modular invariants, are the $\mathcal{A}_k$, $\mathcal{D}_{k\equiv0}$, $\mathcal{E}_{5}$, $\mathcal{E}_{9}$ and $\mathcal{E}_{21}$ graphs. The others are module graphs. The notation $k \equiv 0$ means $k=0 \mod 3$.}
\normalsize
\end{table}

\end{document}